\documentclass[fleqn,usenatbib]{mnras}

\usepackage{newtxtext,newtxmath}
\usepackage{color}

\usepackage[T1]{fontenc}

\DeclareRobustCommand{\VAN}[3]{#2}
\let\VANthebibliography\thebibliography
\def\thebibliography{\DeclareRobustCommand{\VAN}[3]{##3}\VANthebibliography}

\usepackage{graphicx}	
\usepackage{amsmath}	
\usepackage{ulem}
\usepackage{textcase}

\title[$\tauT-f$ degeneracy breaking]{Breaking the $\tauT-f$ degeneracy of the kinetic Sunyaev Zel'dovich cosmology in redshift space}

\author[L. Xiao \& Y. Zheng]{
Liang Xiao,$^{1}$
Yi Zheng$^{1,2,3}$\thanks{E-mail: zhengyi27@mail.sysu.edu.cn}
\\
$^{1}$School of Physics and Astronomy, Sun Yat-sen University, 2 Daxue Road, Tangjia, Zhuhai, 519082, China\\
$^{2}$CSST Science Center for the Guangdong-Hong kong-Macau Greater Bay Area, SYSU, Zhuhai, 519082, China\\
$^{3}$Key Laboratory for Particle Astrophysics and Cosmology (MOE)/Shanghai Key Laboratory for Particle Physics and Cosmology, Shanghai 200240, China\\}

\date{Accepted 2023 July 24. Received 2023 July 23; in original form 2023 May 18}

\pubyear{2023}

\newcommand{\beq}{\begin{equation}}
\newcommand{\eeq}{\end{equation}}
\newcommand{\bea}{\begin{eqnarray}}
\newcommand{\eea}{\end{eqnarray}}
\newcommand{\bi}{\begin{itemize}}
\newcommand{\ei}{\end{itemize}}
\newcommand{\bfi}{\begin{figure}[!t]
\epsfxsize=9cm
\epsffile}
\newcommand{\bfig}{\begin{figure*}[!t]
\center{}
\epsfxsize=15cm
\epsffile}
\newcommand{\efi}{\end{figure}}
\newcommand{\efig}{\end{figure*}}
\newcommand{\no}{\nonumber}
\newcommand{\mpch}{{\rm Mpc}/h}
\newcommand{\hmpc}{h/{\rm Mpc}}
\newcommand{\bfk}{{\bf k}}
\newcommand{\bfv}{{\bf v}}
\newcommand{\tauT}{\tau_{\rm T}}
\newcommand{\tauTf}{\tau_{\rm T}-f}
\newcommand{\bfs}{{\bf s}}
\newcommand{\bfx}{{\bf x}}

\begin{document}
\label{firstpage}
\pagerange{\pageref{firstpage}--\pageref{lastpage}}
\maketitle

\begin{abstract}
The `optical depth-linear growth rate' ($\tauT-f$) degeneracy is a long-standing problem in the kinetic Sunyaev Zel'dovich (kSZ) cosmology. It can be broken in redshift space, where the velocity field leaves its own distinct imprint on the galaxies' redshift space positions and provides valuable information of the linear growth rate. We validate this idea with the Fisher matrix and Monte Carlo Markov Chain techniques in this work, finding that the level of this degeneracy breaking is further enhanced on non-linear scales due to the non-linear evolution of the density and velocity fields, {if we have a good prior knowledge of the non-linear bias of galaxies}. This result emphasizes the importance of the redshift space analysis of the kSZ effect and its potential as a powerful cosmological probe, especially on non-linear scales. As a by-product, we develop a non-linear model of the redshift space density-weighted pairwise kSZ power spectrum. The fitted $f$ and $\tauT$ values from this model are shown to be accurate within $1-2\sigma$ ranges of the fiducial ones when confronted to the mock galaxies mimicking a DESI+CMB-S4 survey combination, even on small scales of $k\sim0.5\hmpc$.
\end{abstract}
\begin{keywords}
large-scale structure of Universe -- intergalactic medium -- cosmic background radiation
\end{keywords}



\section{Introduction}
\label{sec:Intro}

One of the big questions we are still asking in the literature is that why the Universe is accelerating \citep{Acceleration1,Acceleration2}. To answer this question, a variety of theoretical models (modified gravity or dark energy) are developed~\citep{Clifton2012,Brax2018}.  These models are largely degenerated with each other, and in order to break the degeneracy, we intend to simultaneously measure the expansion and structure growth histories of our Universe~\citep{weinbergreview,Joyce2016,Koyama2016}. For the former, cosmic probes like standard candle~\citep{Riess2022}, ruler~\citep{Eisenstein05}, siren~\citep{Abbott2017} and time delay technique~\citep{Wong2020,Treu2022} guarantee an accurate and precise measurement, and for the latter, we resort to the detection of the cosmic peculiar velocity field.

The peculiar velocity of a galaxy is its velocity component that deviates from the Hubble flow. It is induced by the gravitational attraction generated by the small scale inhomogeneity of our Universe. The linear continuity equation directly relates the peculiar velocity field, in particular its divergence, to the matter density growth rate, making it an unbiased tracer of the underlying matter density field. As a result, the peculiar velocity field is an important additional probe for the multi-tracer cosmological analysis~\citep{Koda13,Howlett17,Munchmeyer2019} and a unique cosmic probe which helps understand the fundamental physics such as inflation~\citep{Munchmeyer2019}, dark energy~\citep{Okumura22}, modified gravity~\citep{Roncarelli2018,PanZ2019,Mitchell2021}, massive neutrino \citep{Mueller2015,Whitford2022,ZhouSR2022} and so on. 

The current peculiar velocity detection methodologies can be divided into three categories. The first is through the redshift space distortion (RSD) effect~\citep{Jackson72,Kaiser87}. This method indirectly detects the velocity field  by its anisotropic imprint on the galaxy clustering pattern in redshift space. Having the highest detection significance within the three, the RSD effect has been successfully utilized by various galaxy surveys to measure the structure growth history and constrain cosmological models (e.g., \citealt{Peacock01,Guzzo08,Samushia12,Alam2017,Hector2020}). The second kind of methods depend on extra measurements that reveal the real space distance of a galaxy, which will then be subtracted from the measured redshift space distance  and unveil its peculiar velocity. This extra distance measure can come from Tully-fisher~\citep{Tully1977} or Fundamental plane relationship~\citep{Djorgovski1987}, Type Ia supernova~\citep{Zhang08a}, Gravitational wave event~\citep{Abbott2017} and so on. Although their utility is limited either by large intrinsic uncertainties or limited detection events, the next generation surveys will make improvements on these downsides and make the so-called peculiar velocity surveys good complements to the RSD effect~\citep{Howlett17,Kim2020,Adams2020}. Lastly,  the third method is the kinetic Sunyaev Zel'dovich (kSZ) effect that we will discuss in this paper.

The kSZ effect is a secondary CMB anisotropy in which the temperature of CMB photons is slightly changed by their inverse Compton scattering off of free electrons with a bulk motion~\citep{kSZ1970,kSZ1972,kSZ1980}. It detects the momentum field of free electrons in our Universe, as illustrated by 
\beq
\label{eq:deltaT_kSZ}
\frac{\delta T_{\rm kSZ}(\hat{n})}{T_0} = -\sigma_{\rm T}\int dl n_{\rm e} \left(\frac{{\bfv}_{\rm e}\cdot \hat{n}}{c}\right)  \,.
\eeq
Here $T_0\simeq 2.7255\rm K$ is the averaged CMB temperature, $\sigma_{\rm T}$ is the Thomson-scattering cross-section, $c$ is the speed of light, $\hat{n}$ is the unit vector along the line of sight (LOS), $n_{\rm e}$ is the physical free electron number density, $\bfv_{\rm e}$ is the proper peculiar velocity of free electrons, defined to be positive for those recessional objects, and the integration $\int dl$ is along the LOS given by $\hat{n}$.

As a result, cosmological applications of the kSZ effect are two-fold. First it measures the (ionized) baryon distribution in our Universe and can be used to constrain the Epoch-of-Reionization (EoR) history~\citep{Alvarez2016,ChenN2023} or the gas distribution within halos~\citep{Schaan16,Sugiyama18} and filaments~\citep{ZhengY2023}. In particular, since $\delta T_{\rm kSZ}(\hat{n})$ is linearly proportional to $n_{\rm e}$ and independent of the gas temperature, it is well suited to search for missing baryons \citep{ZhengY2023,Carlos2015,Shao2016,Lim2020,Jonas2021} which are believed to mainly reside in environments {which have a intermediate density and temperature between the intra-cluster medium and cooler photoionized baryons that can be picked up through their residual HI absorption}. Second, it directly measures the LOS peculiar velocity field, which in turn helps constrain different aspects of the cosmological model~\citep{Okumura22}, e.g., Copernican principle~\citep{Zhang11b}, primordial non-Gaussianity~\citep{Munchmeyer2019,Kumar2022}, dark energy~\citep{Pen2014}, modified gravity~\citep{Roncarelli2018,Zheng20,Mitchell2021}, massive neutrino~\citep{Roncarelli2017} and so on.

We focus on the late-time kSZ effect in this work, which denotes the kSZ effect after the EoR. It can be probed via cross-correlations between CMB data and tracers of large-scale structure at different redshifts, a technique known as the kSZ tomography. Since its first measurement one decade ago~\citep{Hand12}, the current kSZ detection significance by adopting this technique is around $4\sim7\sigma$, with combinations of different CMB data and galaxy catalogs, e.g., SPT+DES~\citep{Soergel16}, ACT+SDSS~\citep{Schaan2021,Calafut2021}, Planck/WMAP+unWISE~\citep{Kusiak2021}, Planck+DESI(photo-z)~\citep{Chen2022}, et al. With the next generation of surveys, 
signal-to-noises (S/N's) of such measurements can reach $\mathcal{O}(100)$ and beyond~\citep{Sugiyama18,Smith2018}.

The major contribution of the detected kSZ signal by kSZ tomography comes from free electrons in and around dark matter halos, and within the foreground galaxy catalog, most CMB photons are assumed to encounter only one large halo during their journey to us. In this case, Eq.~(\ref{eq:deltaT_kSZ}) can be reformed as 
\beq
\label{eq:deltaT_kSZ2}
\frac{\delta T_{\rm kSZ}(\hat{n}_i)}{T_0} = -\frac{\tau_{{\rm T},i}}{c}\bfv_i\cdot\hat{n}_i\approx-\frac{\tauT}{c}\bfv_i\cdot\hat{n}_i\,,
\eeq
where $\tau_{{\rm T},i}=\int dl\sigma_{\rm T}n_{{\rm e},i}$ is the optical depth of the $i$th halo and $\tauT$ is the average optical depth of a halo sample. Eq.~(\ref{eq:deltaT_kSZ2}) shows that, $\tauT$ fully couples with the galaxy velocity field and modulates its overall amplitude in the kSZ measurement. As a result, in a real space kSZ observation it is difficult to decouple $\tauT$ from those cosmological parameters controlling the velocity field amplitude, such as the linear growth rate $f\equiv d\ln D/d\ln a$ where $D$ is the linear growth factor and $a$ is the scale factor. It leads to the so-called $\tauT-f$ degeneracy in the kSZ cosmology and a fact that, little cosmological information can be extracted solely from the kSZ observations without a concrete knowledge of the optical depth. This degeneracy is rigorous on linear scales in real space, as we prove in Appendix~\ref{app:degeneracy_proof} using the continuity equation.

This degeneracy can be illustrated in alternative ways. For example, in some works the $\tauT$ parameter is alternatively expressed as a constant velocity bias $\tilde{b}_v$ in the linearly reconstructed radial velocity field $\tilde{v}_{\hat{n}}$ via the kSZ tomography~\citep{Smith2018},
\beq
\label{eq:vel_recons}
\tilde{v}_{\hat{n}}(\bfk) = \mu\frac{\tilde{b}_vafH}{k} \delta(\bfk)+({noise})\,.
\eeq
Here $\tilde{v}_{\hat{n}}\equiv\tilde{\bfv}\cdot\hat{n}$ and $\mu\equiv\cos(\hat{\bfk}\cdot\hat{n})$ denotes the cosine of the angle between $\bfk$ and the LOS. In the cosmological inference, $b_v$ is then marginalized and loosens the constraints on parameters relative to the amplitude of the velocity power spectrum like $f$.

Several ideas were proposed to break this degeneracy and enhance the scientific return of the kSZ cosmology: (1) it was proposed in~\cite{Sugiyama2016} that the synergy between kSZ effect and redshift space distortion (RSD) effect can break this degeneracy; (2) a comprehensive model \citep{Flender2017} or empirical relationship \citep{David2016,Battaglia2016,Soergel2018} was constructed to predict the value of $\tauT$; (3) extra observations like fast radio burst can directly measure the optical depth \citep{Madhavacheril2019}; (4) studying the three-point statistic of the pairwise velocity field can avoid the $\tauT$ dependency in a cosmological inference~\citep{Kuruvilla2022}; and so on. 

Being innovative and promising though, all of these ideas except method (4) rely on extra information of $\tauT$ or $f$ from other observations. {On the contrary, we explore in this work the possibility of breaking this degeneracy by the non-linear kSZ effect in redshift space, and  
the idea is simple.} Since we actually apply the kSZ tomography technique in redshift space where the velocity field leaves its own imprint on the galaxies' redshift space positions, the extra information of $f$ can be provided by the RSD effect encoded in the observed kSZ effect itself, and the $\tauT-f$ degeneracy can be robustly broken. 

In the following sections, we will first use Fisher matrix technique to reveal the severe $\tauT-f$ degeneracy on linear scales, and by combining the dipole and octopole of the pairwise kSZ power spectrum in redshift space, the degeneracy can be broken in a limited level. Then, by developing non-linear kSZ models, we extend the Fisher matrix analysis to non-linear scales and find that the degeneracy can be broken in a dramatic level therein. This is the main result of this paper, and it is made more solid by being verified alternatively with a Monte Carlo Markov Chain (MCMC) analysis on mock galaxy catalogs from simulations. {After this, we discuss the impact of the unknown non-linear density bias on our results and find that it heavily weakens the level of the degeneracy breaking, unless a tight prior on the non-linear bias is given by extra data like galaxy clustering or weak lensing. In summary, this work will show how $\tauT$ and $f$ can be simultaneously constrained with the kSZ effect alone if we work on non-linear scales in redshift space, and it naturally benefits the cosmological applications of this effect in the future. Although the robustness of the conclusion is weakened by the non-linear galaxy density bias, it still highlights the importance of studying the kSZ effect on non-linear scales, including the development of an accuracy non-linear kSZ model.}



The rest of this paper will be organized as follows. In Sec.~\ref{sec:models} we present the developed theoretical kSZ models. In Sec.~\ref{sec:mock} we introduce the mock galaxy catalogs constructed from simulations. In Sec.~\ref{sec:method} we present the methodologies adopted in our work, including the measurements of the pairwise kSZ power spectra from the mock catalogs, the Fisher matrix analysis and the MCMC analysis. In Sec.~\ref{sec:result} we show the results and Sec.~\ref{sec:discussion} is dedicated to conclusion and discussion. 

\section{Models}
\label{sec:models}

The kSZ tomography technique can be implemented with various estimators. We refer readers to~\cite{Smith2018} for a good summary. Although in this work the idea of the degeneracy breaking is justified with the widely adopted \textit{pairwise kSZ estimator}, the main conclusion will also hold for others.

Biased tracers, such as galaxies living in dark matter halos containing free electrons, tend to move toward each other due to the mutual gravitational attraction. This peculiar kinematic pattern leaves a distinct feature in the CMB map, which can be captured by the pairwise kSZ estimator~\citep{Hand12}. With this estimator,  the density weighted pairwise kSZ correlation function in redshift space and its Fourier counterpart can be expressed as~\citep{Sugiyama2017}
\beq
\xi^{\rm s}_{\rm kSZ}(\bfs) = \left\langle -\frac{V}{N^2} \sum_{i,j}\left[\delta T_{\rm kSZ}(\hat{n}_i)-\delta T_{\rm kSZ}(\hat{n}_j)\right]\delta_{\rm D}({\bfs}-{\bfs}_{ij}) \right\rangle \,,
\label{eq:xi_kSZ}
\eeq
\beq
P^{\rm s}_{\rm kSZ}(\bfk) = \left\langle -\frac{V}{N^2} \sum_{i,j}\left[\delta T_{\rm kSZ}(\hat{n}_i)-\delta T_{\rm kSZ}(\hat{n}_j)\right]e^{-i \bfk\cdot {\bfs}_{ij}} \right\rangle \,,
\label{eq:P_kSZ}
\eeq
where $V$ is the survey volume, $N$ is the number of galaxies, and ${\bfs}_{ij}={\bfs}_i-{\bfs}_j$ is the pair separation vector in redshift space. We have assumed here that $\tauT$ is not correlated with the galaxy density and velocity fields.

Since the $\tauT-f$ degeneracy and the way it is broken are more transparent in Fourier space, we will work with power spectrum from now on. Inserting Eq.~(\ref{eq:deltaT_kSZ2}) into Eq.~(\ref{eq:P_kSZ}), we have
\bea
P^{\rm s}_{\rm kSZ}(\bfk) &\simeq& \left(\frac{T_0\tauT}{c}\right)\left\langle \frac{V}{N^2} \sum_{i,j}\left[\bfv_i\cdot\hat{n}_i-\bfv_j\cdot\hat{n}_j\right]e^{-i \bfk\cdot {\bfs}_{ij}} \right\rangle \, \no \\
&\equiv& \left(\frac{T_0\tauT}{c}\right)P^{\rm s}_{\rm pv}(\bfk)\,.
\label{eq:P_pv}
\eea
Here $P^{\rm s}_{\rm pv}$ is the density-weighted pairwise LOS velocity power spectrum in redshift space. 

It is proven in \cite{Sugiyama2016,Sugiyama2017} that in redshift space, $P^{\rm s}_{\rm pv}$ is related to the density power spectrum $P^{\rm s}$ by a simple relation
\beq
P^{\rm s}_{\rm pv}(\bfk) =\left(i\frac{aHf}{k\mu}\right) \frac{\partial}{\partial f} P^{\rm s}(\bfk)\,,
\label{eq:P_kSZ_der}
\eeq
where $a$ is the scale factor, $H$ is the Hubble parameter at redshift $z$. This relation holds for any object such as dark matter particles, halos, galaxies, et al. Since observationally we measure the kSZ effect in redshift space, our kSZ models will be based on this formula.

The detailed derivation of this relationship, including its extensions to higher order moments of the density-weighted velocity field, can be found in~\cite{Sugiyama2016,Sugiyama2017}. In Appendix~\ref{app:eq_ksz_der_proof} we provide a simple proof of Eq.~(\ref{eq:P_kSZ_der}) itself.

\subsection{Linear model}
\label{subsec:linear_model}
\begin{figure}
\begin{center}
\includegraphics[width=\columnwidth]{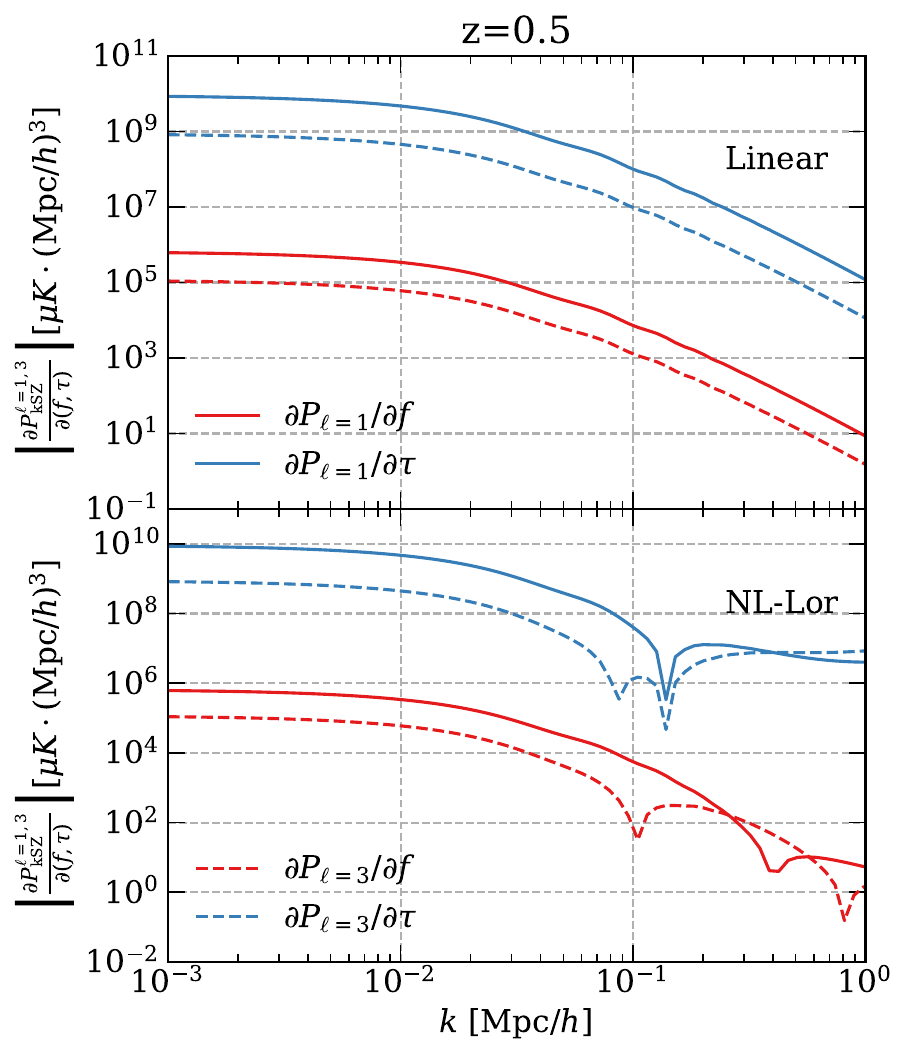}
\end{center}
\caption{The absolute values of the derivatives of $P^{\ell=1,3}_{\rm kSZ}$ with respect to $\tauT$ and $f$ calculated by the linear model ({\it top panel}) and the non-linear `NL-Lor' kSZ model ({\it bottom panel}). {The linear bias $b_1$ is adopted in the top-panel calculations and $b(k)=b_1+b_2k^2$ is adopted in the bottom-panel calculations.}}
\label{fig:derivative}
\end{figure}

First we derive the linear model of $P^{\rm s}_{\rm kSZ}$. Inserting the linear RSD model of $P^{\rm s}$~\citep{Kaiser87}
\beq
P^{\rm s}(k,\mu) = \left(b_1+f\mu^2\right)^2P_{\rm lin}(k)
\eeq
into Eqs.~(\ref{eq:P_pv}) and ~(\ref{eq:P_kSZ_der}), we obtain
\beq
\label{eq:P_kSZ_lin}
P^{\rm s,lin}_{\rm kSZ}(k,\mu)=\left(\frac{T_0\tauT}{c}\right)2iaHf\mu(b_1+f\mu^2)\frac{P_{\rm lin}(k)}{k}\,.
\eeq
Here $b_1$ is the linear galaxy bias, $P_{\rm lin}(k)$ is the linear dark matter density power spectrum at the redshift $z$. The multipoles of a power spectrum are defined as
\beq
\label{eq:multipoles}
P_\ell(k) = \frac{2\ell+1}{2}\int^1_{-1}d\mu P(k,\mu){\cal L}_\ell\,,
\eeq
in which ${\cal L}_\ell$ is the Legendre polynomial. Then the dipole and the octopole of $P^{\rm s,lin}_{\rm kSZ}$ are
\bea
\label{eq:P_kSZ_lin_l1}
P^{\ell=1,\rm lin}_{\rm kSZ}&=&2iaH\left(\frac{T_0}{c}\right)\left(b_1f\tauT+\frac{3}{5}f^2\tauT\right)\frac{P_{\rm lin}}{k}\,, \\
\label{eq:P_kSZ_lin_l3}
P^{\ell=3,\rm lin}_{\rm kSZ}&=&2iaH\left(\frac{T_0}{c}\right)\left(\frac{2}{5}f^2\tauT\right)\frac{P_{\rm lin}}{k}\,. 
\eea

As shown in Eq.~(\ref{eq:P_kSZ_lin}), on linear scales $P^{\rm s}_{\rm kSZ}$ can be decomposed into two terms of $\tauT$ and $f$ ($f\tauT\mu$ and $f^2\tauT\mu^3$) with different $\mu$ dependence. Likewise, the dipole in Eq.~(\ref{eq:P_kSZ_lin_l1}) contains terms of $f\tauT$ and $f^2\tauT$ and the octopole in Eq.~(\ref{eq:P_kSZ_lin_l3}) only contains $f^2\tauT$. This anisotropic property in the redshift space kSZ power spectrum induced by the RSD effect can thus spontaneously break the $\tauT-f$ degeneracy and the main idea of this paper is illustrated on linear scales. 

With this qualitative demonstration in mind, the quantitative level of the degeneracy breaking is limited on linear scales. As an example, $b_{\rm L}$ is usually larger than 1 in Eq.~(\ref{eq:P_kSZ_lin}) and $f\mu^2$ is always smaller than 1, the term  $\propto b_{\rm L}f\tauT$ thus dominates the linear power spectrum and diminishes the level of the degeneracy breaking on linear scales. This situation is better illustrated on the top panel of Fig.~\ref{fig:derivative}, where we plot the absolute values of the derivatives $\partial P^{\ell=1,3}_{\rm kSZ}/\partial(f,\tauT)$ of the linear model. We see that the ways in which the linear $P^{\ell=1,3}_{\rm kSZ}$ are sensitive to $f$ and $\tauT$ are quite similar to each other at all scales, demonstrating the difficulty of the degeneracy breaking on linear scales, or if we wrongly apply the linear model on non-linear scales.

By the same argument, we notice that the linear bias $b_1$ is also degenerated with $f$ and $\tauT$ in Eq.~(\ref{eq:P_kSZ_lin}). {This degeneracy also holds true at non-linear scales, as we can see in the next section. Observational determination of the density bias requires extra observational data such as galaxy clustering and weak lensing. In the following discussions, we will first assume an {\it a priori} known bias term directly measured from simulations, which enables us to focus on the degeneracy breaking between $\tauT$ and $f$. Then we will incorporate two non-linear bias parameters in the MCMC fitting, and discuss how the non-linear bias term weakens the parameter breaking in a kSZ data analysis alone.} 
Furthermore, we assume a unity velocity bias $b_v\equiv P_{\delta\theta}/P_{\delta\delta}=1$ in this work~\citep{Zheng14b,Junde18vb}, where $\theta\equiv-\nabla\cdot\bfv/afH$.

\subsection{Non-linear models}
\label{subsec:nonlinear_models}

The situation is quite different on non-linear scales where the linear kSZ model does not work. For deriving the non-linear pairwise kSZ model, we start from a phenomenological RSD model, with a Kaiser term and a Finger-of-God (FoG) term~\citep{Scoccimarro04,Zhangrsd}:
\beq
P^{\rm s}(k,\mu) = \left(b+fW\mu^2\right)^2P_{\delta\delta}(k)D^{\rm FoG}(k\mu\sigma_v)\,.
\label{eq:rsd}
\eeq
Here $b=b(k)$ is the non-linear galaxy bias, $P_{\delta\delta}(k)$ is the non-linear dark matter density power spectrum at redshift $z$, and $W(k)\equiv P_{\delta\theta}/P_{\delta\delta}$ quantifies the non-linear correlation between density and velocity fields~\citep{Zhangrsd}. 

Motivated by the perturbation theory and verified with N-body simulations, $W(k)$ can be described by the following fitting function~\citep{Zheng13}
\beq
\label{eq:W}
W(k) = \frac{1}{1+\Delta\alpha(z)\Delta^2_{\delta\delta}(k,z)}\,,
\eeq
where $\Delta^2_{\delta\delta}(k,z) = k^3P_{\delta\delta}/2\pi^2$ is the dimensionless density power spectrum, and the redshift dependence of $\Delta\alpha$ can be fitted well by a linear relation~\citep{Zheng13}
\beq
\label{eq:Delta_alpha}
\Delta\alpha(z) = \Delta\alpha(z=0)+0.082z = 0.331+0.082z\,.
\eeq

In order to compensate the inaccuracy of our simple Kaiser term modelling, we allow the functional form of $D^{\rm FoG}$ to vary. We try Gaussian and Lorentzian $D^{\rm FoG}$, namely
\begin{eqnarray}
\label{eq:FoG}
	D^{\rm FoG}(k\mu\sigma_v)=
	\left\{
	\begin{array}{cc}
		\exp(-k^2\mu^2\sigma_v^2/H^2)\,\,\,\, {\rm Gaussian}\,,  \\
		(1+k^2\mu^2\sigma_v^2/H^2)^{-1}\,\,\,\, {\rm Lorentzian}\,.
	\end{array}
	\right.
\end{eqnarray}
Here $\sigma_v^2$ is the LOS velocity dispersion and $\sigma_v^2=\int f^2H^2P_{\rm lin}(k)dk/6\pi^2$ in linear theory. It will be treated as a free parameter in the model fitting and absorbs systematic errors originating from the model inaccuracies.

By substituting Eqs.~(\ref{eq:rsd}) and~(\ref{eq:FoG}) into Eq.~(\ref{eq:P_kSZ_der}), we derive the corresponding formula of $P^{\rm s}_{\rm kSZ}(k,\mu)$~\footnote{When making the derivative, we adopt this expression for $\sigma_v$: $\sigma_v^2=\int f^2H^2P_{\delta\delta}(k)dk/6\pi^2$}. For Gaussian $D^{\rm FoG}$~\citep{Zheng20},
\bea
\label{eq:P_kSZ_Gau}
P^{\rm s,Gau}_{\rm kSZ}(k,\mu)&=&\left(\frac{T_0\tauT}{c}\right)2iaHf\mu(b+fW\mu^2)\frac{P_{\delta\delta}(k)}{k} \no \\
&&\times  S_{\rm G}(k,\mu) \exp(-k^2\mu^2\sigma_v^2/H^2)\,, \\
S_{\rm G}(k,\mu)&=&W-(b/f+W\mu^2)k^2\sigma_v^2/H^2\,.\no
\eea
For Lorentzian $D^{\rm FoG}$,
\bea
\label{eq:P_kSZ_Lor}
P^{\rm s,Lor}_{\rm kSZ}(k,\mu)&=&\left(\frac{T_0\tauT}{c}\right)2iaHf\mu(b+fW\mu^2)\frac{P_{\delta\delta}(k)}{k} \no \\
&&\times  S_{\rm L}(k,\mu) \frac{1}{1+k^2\mu^2\sigma_v^2/H^2}\,, \\
S_{\rm L}(k,\mu)&=&W-(b/f+W\mu^2)\frac{k^2\sigma_v^2/H^2}{1+k^2\mu^2\sigma_v^2/H^2}\,.\no
\eea
As pointed out in~\cite{Zheng20}, the shape kernel $S(k,\mu)$ is a consequence of the competition between the Kaiser and FoG effects. As $k$ increases, $S(k,\mu)$ decreases from 1 to $-\infty$ and induces the turnover of the sign of $P^{\ell=1}_{\rm kSZ}$ at quasi-linear scales as shown on the top panel of Fig.~\ref{fig:Pkl1l3}.

Depending on which FoG term we choose, and if we set $W=1$ or to be Eq.~(\ref{eq:W}), the non-linear models can be divided into 4 categories in this work, and they are labelled as `NL-Gau', `NL-Gau-W', `NL-Lor', and `NL-Lor-W' separately.

Compared to the linear model, the $k$ and $\mu$ dependence of $P^{\rm s}_{\rm kSZ}$ in non-linear models are much more complicated. On the bottom panel of Fig.~\ref{fig:derivative}, we display the absolute values of the derivatives $\partial P^{\ell=1,3}_{\rm kSZ}/\partial(f,\tauT)$ of the NL-Lor model. Four derivatives are different from each other at $0.1\hmpc<k<1\hmpc$, and these differences will enable an easier separation of $f$ and $\tauT$ when we work on non-linear scales. This phenomenon further drives the main idea of the $\tauT-f$ degeneracy breaking in this paper and we will show how it works in Sec.~\ref{sec:result}.


\section{Mock catalogs}
\label{sec:mock}

\begin{table*}
\begin{tabular}{cccccccccc}
\hline\hline
 &FWHM & Noise & Galaxy & Redshift & $V$ & $\bar{n}_{\rm g}$  & $M_{\rm avg}$ & $\tauT$ \\
&$[{\rm arcmin}]$ & $[\mu K\mathchar`-{\rm arcmin}]$ &  &  & $[(h^{-1}\,{\rm Gpc})^3]$ & $[(h^{-1}\,{\rm Mpc})^{-3}]$ &  $[h^{-1}\, M_{\odot}]$ &\\
\hline
CMB-S4 + BOSS & $1$ & $2$ & LRG & 0.5 $(0.43<z<0.70)$ & 4.0 & $3.2\times10^{-4}$ & $2.6\times10^{13}$ & $4.5\times10^{-5}$\\
CMB-S4 + DESI & $1$ & $2$ & LRG & 0.8 $(0.65<z<0.95)$ & 13.5 & $2\times10^{-4}$ & $2.6\times10^{13}$ & $6\times10^{-5}$\\
\hline\hline
\end{tabular}
\caption{The assumed survey specifications. The first two columns show beam size and detector noise in a CMB-S4-like CMB experiment. We assume an ideal beam size of 1 arcmin. The third to eighth column individually shows the type of galaxies, redshift $z$, comoving survey volume $V$, mean comoving number density $\bar{n}_{\rm g}$, the averaged halo mass $M_{\rm avg}$ measured from the simulation data and the average optical depth $\tauT$ of the galaxy catalog calculated assumed a Gaussian projected gas profile~\citep{Zheng20}.}
\label{table:survey}
\end{table*}
\begin{figure}
\begin{center}
\includegraphics[width=\columnwidth]{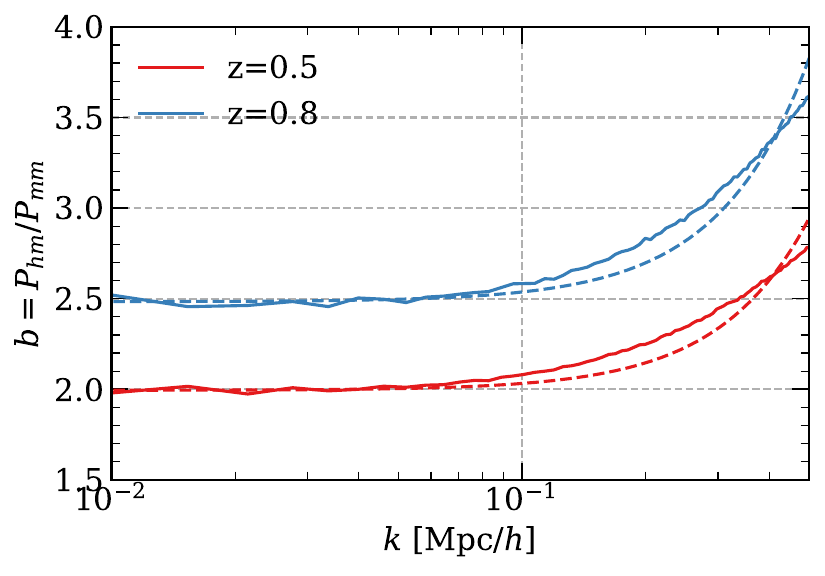}
\end{center}
\caption{The measured galaxy density bias $b(k)=P_{\rm gm}/P_{\rm mm}$ from simulations, shown in solid lines. {The dashed lines show the theoretical $b(k)=b_1+b_2k^2$ (Eq.~(\ref{eq:NL_bk})).} }
\label{fig:bias}
\end{figure}

In the following sections we will compare our models to mock galaxy catalogs and reveal the $\tauT-f$ degeneracy breaking. The mocks are constructed from the GR part of the ELEPHANT simulation suite~\citep{Cautun2018} run by the \textsc{ecosmog} code \citep{ECOSMOG,Bose2017}. The simulation has a box size of $1024\mpch$ and a dark matter particle number of $1024^3$. There are in total five independent realizations run under the $\rm \Lambda CDM$ model, and the background expansion is quantified by the WMAP9 cosmology \citep{WMAP9}, $\{\Omega_{\rm b}, \Omega_{\rm CDM}, h, n_s, \sigma_8 \} = \{0.046, 0.235, 0.697, 0.971, 0.82\}$.

We analyze two mock catalogs from simulation snapshots of $z=0.5$ and $z=0.8$ in this work. They are constructed to  separately mimic the CMASS sample of the BOSS survey \citep{Manera2013} and the LRG galaxy sample of the DESI survey between $z= 0.65$ and $z=0.95$ \citep{DESI16I, Sugiyama18}. The methodology of generating the galaxy mocks are as follows.  

At $z=0.5$, the mock galaxy catalogs are generated by applying  the HOD model suggested in \cite{ZhengZheng2007} to the dark matter halo catalogs generated  by \textsc{rockstar}~\citep{Rockstar}. The best-fitted HOD parameters are from the CMASS data \citep{Manera2013}.  At $z=0.8$, we simply select all dark matter halos with $M>10^{13}M_\odot/h$ to represent the LRG galaxies. In Table~\ref{table:survey} we list some details of the mock catalogs, and a more thorough description of the simulations and catalogs can be found in \cite{Cautun2018,Cesar2019,Alam2021}.

In Fig.~\ref{fig:bias} we show the measured density biases $b(k)$ of the constructed mocks, evaluated by the ratio between the measured galaxy-matter cross-power spectrum and the matter-matter auto-power spectrum, $b(k)=P_{\rm gm}/P_{\rm mm}$. The bias measured by this estimator deviates from the true $b(k)$ by a correlation coefficient $\eta \le 1$ between the galaxy and matter density fields. This coefficient tends to be unity at linear scales, and  its impact at small scales will be partially accounted for by the nuisance parameter $\sigma_v$ in our model fitting. 

{In a realistic data analysis, the potential systematic errors coming from $\eta$ will be avoided by a robust model of $b(k)$~\citep{McDonald_bias,Hector_bias,Beutler_bias,Zheng16c}. Here we adopt a simple yet robust phenomenological non-linear bias for $b(k)$ as~\citep{Desjacques18biasreview}
\beq
\label{eq:NL_bk}
b(k) = b_1+b_2k^2\,.
\eeq
We estimated the linear density bias $b_1$ by averaging $b(k)$ at $k<0.06\hmpc$, and the measured $b_{1,\rm fid} =\{2.00,2.48\}$ at $z=\{0.5,0.8\}$. We fit the non-linear bias parameter $b_2$ with the measured $b(k)$ at $k\le0.5\hmpc$, and the fitted $b_{2,\rm fid} = \{3.80,5.37\}$ at $z=\{0.5,0.8\}$. The theoretical $b(k)$'s are shown with dashed lines in Fig.~\ref{fig:bias}. The fitted $b_{1,\rm fid}$ and $b_{2,\rm fid}$ will be used as the fiducial values of $b_1$ and $b_2$ in the following Fisher matrix and MCMC analysis.}

\section{Methodology}
\label{sec:method}
In this section we present the methodologies we apply for calculating $P^\ell_{\rm kSZ}(k,\mu)$ from simulations and for forecasting the model parameter constraints in this work.

\subsection{Pairwise kSZ multipoles}
\begin{figure}
\begin{center}
\includegraphics[width=\columnwidth]{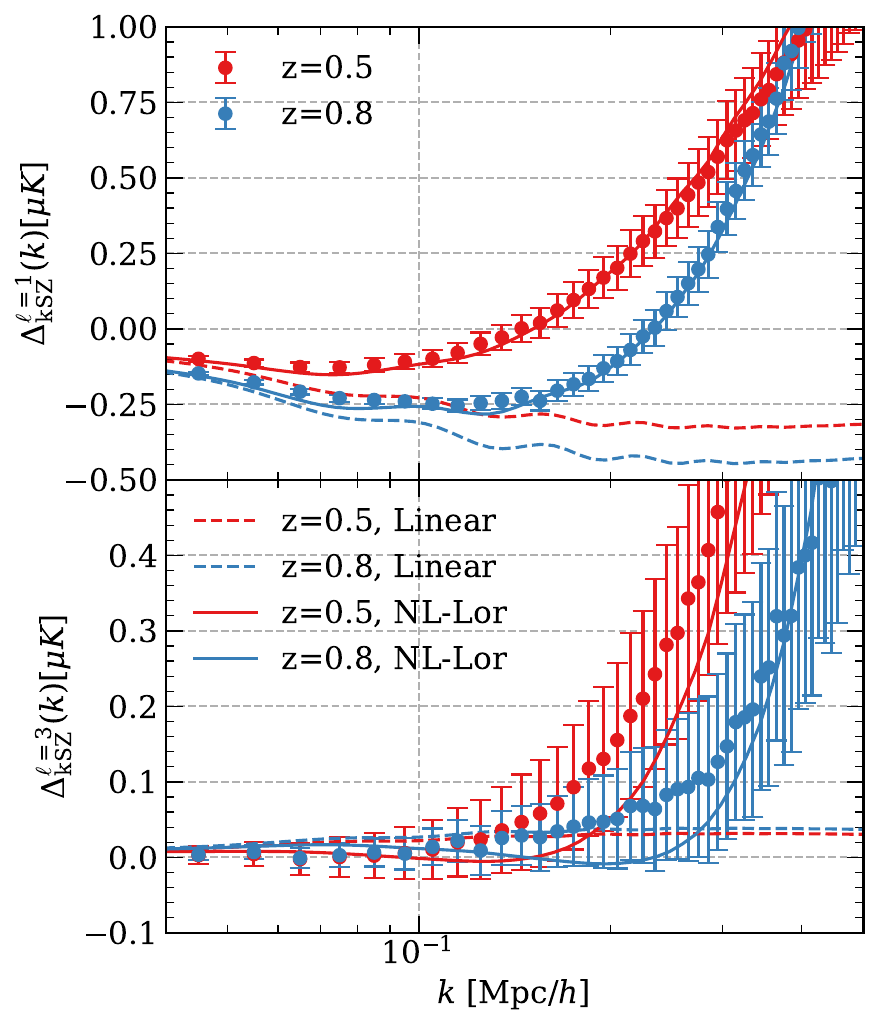}
\end{center}
\caption{The pairwise kSZ multipoles with error bars. $\Delta_{\rm kSZ}^{\ell=1,3}(k)\equiv i^\ell k^3P_{\rm kSZ}^{\ell=1,3}/(2\pi^2)$. The dashed lines represent the linear model and the solid lines show the NL-Lor model. The measured $b(k)$ terms are adopted.}
\label{fig:Pkl1l3}
\end{figure}

As the first step, we move the mock galaxies from their real to redshift space positions with the following formula:
\beq
\label{eq:mapping}
{\bf s}={\bf r}+\frac{{\bf v} \cdot \hat{n}}{a(z)H(z)}\hat{n}\,,
\eeq
where {\bf r} is the real space position and {\bf s} is the redshift space position.

For the computation of $P^{\rm s}_{\rm pv}(k,\mu)$, we adopt a field-based estimator which is equivalent to the particle-based one in Eq.~(\ref{eq:P_pv})~\citep{Sugiyama2016},
\beq
(2\pi)^3\delta_{\rm D}(\bfk+\bfk')P^{\rm s}_{\rm pv}(\bfk)=\left<p_{\rm s}(\bfk)\delta_{\rm s}(\bfk')-\delta_{\rm s}(\bfk)p_{\rm s}(\bfk')\right> \,,
\label{eq:P_pv_field}
\eeq
where $p_{\rm s}(\bfk)$ and $\delta_{\rm s}(\bfk)$ are Fourier counterparts of the radial momentum field $p_{\rm s}({\bfs})=[1+\delta_{\rm s}({\bfs})][\bfv({\bfs})\cdot \hat{n}]$ and the density field $\delta_{\rm s}({\bfs})$ respectively. We sample the $p_{\rm s}(\bfs)$ and $\delta_{\rm s}(\bfs)$ fields on $1024^3$ regular grids using the nearest-grid-point (NGP) method and calculate $p_{\rm s}(\bfk)$ and $\delta_{\rm s}(\bfk)$ fields by the fast Fourier transform (FFT). Eq.~(\ref{eq:P_pv_field}) denotes that the $P^{\rm s}_{\rm pv}$ measurement is actually measuring the cross-correlation between the momentum and density fields.

The measured optical depth of a halo depends on its projected gas density profile and the adopted aperture photometry (AP) filter. Following \cite{Sugiyama18}, we assume a Gaussian gas density profile, and choose an AP filter radius $\theta_{\rm c}$ that can maximize the kSZ detection S/N. We choose a CMB-S4-like survey as our baseline CMB survey~\citep{CMBS42019}, and assume that it has a full overlap between the sky coverage of BOSS and DESI. The chosen specification of the CMB-S4-like survey is shown in Table~\ref{table:survey}. We refer readers to~\cite{Sugiyama18} (Eq. (17) therein) and the appendix of~\cite{Zheng20} for the details of the $\tauT$ calculation. As a result, the estimated $\tauT$'s of two galaxy mocks are separately $\tauT = 4.5\times10^{-5}$ at $z=0.5$ with $\theta_{\rm c} = 1.37'$, and  $\tauT = 6.0\times10^{-5}$ at $z=0.8$ with $\theta_{\rm c} = 1.12'$. They will be set to be the fiducial values of $\tauT$ in the following analysis.  Since $\tauT$ only affects the overall amplitude of $P^{\rm s}_{\rm kSZ}$, we do not expect that the uncertainty of its estimation will affect the conclusion of this paper.

We also calculate the covariance matrix of $P^{\ell=1,3}_{\rm kSZ}(k)$ in a theoretical way. The detailed methodology is presented in  Appendix~\ref{app:covariance}. Although we only consider the Gaussian part of the $P^{\rm \ell=1,3}_{\rm kSZ}$ covariance in our analysis, we find in Appendix~\ref{app:covariance} that this theoretical covariance coincides with the measured one from 100 N-body simulations and is pretty accurate even at scales reaching $k\sim0.5\hmpc$.  Therefore we do not  expect that the simplification in the covariance matrix calculation will significantly bias our results in this work. 

In Fig.~\ref{fig:Pkl1l3}, we plot the measured pairwise kSZ multipoles, $\Delta^{\ell=1,3}_{\rm kSZ}\equiv i^\ell k^3P^{\ell=1,3}_{\rm kSZ}/(2\pi^2)$ (hereafter we will use $P^{\ell}_{\rm kSZ}$ and $\Delta^{\ell}_{\rm kSZ}$ interchangeably to refer to the kSZ power spectrum multipole). On the top panel, the shape kernel $S(k,\mu)$ controls the turnover of $\Delta^{\ell=1}_{\rm kSZ}$ from being negative to positive at quasi-linear scales, showing the significant impact of the FoG effect in the non-linear kSZ power spectrum. For both $\Delta^{\ell=1}_{\rm kSZ}$ and $\Delta^{\ell=3}_{\rm kSZ}$, the linear model deviates from the measurement at very large scales of $k\sim0.05\hmpc$, while our non-linear model can account for the non-linear evolution of the measurement in a much better way.


\subsection{Fisher matrix}
\label{subsec:fisher}
We adopt the fisher matrix technique to predict the constraints that future surveys will impose on the cosmological parameters around a fiducial cosmology. The fisher matrix is calculated by
\beq
\label{eq:fisher_m}
F_{ij} = \frac{\partial X^{\rm T}}{\partial \theta_i}C^{-1}\frac{\partial X}{\partial \theta_j}+(prior)\,.
\eeq 
Given the vector of model parameters $\Theta = \{\theta_i\}$, $X=X(\Theta)$ is the vector of the measured quantities. $C$ is the covariance matrix assumed to be independent of $\Theta$. The \textit{prior} term is usually a diagonal matrix with its diagonal element being $1/\sigma_i^2$, in which $\sigma_i^2$ is the variance of the Gaussian prior of the $i$th parameter. Although we will apply flat priors $[a_i,a_i+b_i]$ in the MCMC analysis,  we practically apply a Guassian prior with $(b_i/2)^2$ being its effective variance in the Fisher matrix calculation. 

In an ideal case of the Gaussian likelihood, $\Delta \theta_i\ge(F^{-1})^{1/2}_{ii}$ according to the Cram\'er–Rao inequality. Therefore we can place a firm lower limit on the parameter error bar $\Delta \theta_i$ that can be attained from surveys. We will quote this lower limit as the 1-$\sigma$ parameter constraint forecast hereafter.

In the following calculations of $F_{ij}$, we choose $\Theta=\{f,\log_{10}\tauT,\sigma_v\}$ with $f \in [0,1]$, $\log_{10} \tauT \in [-7, -3]$ and $\sigma_v \in [0,1000]$. $X(\Theta)$ is chosen to be $P^{\ell=1,3}_{\rm kSZ}(k)$ at $0.035\hmpc\le k\le k_{\rm max}$ and $k_{\rm max}$ is a certain cut-off scale we can vary. $P^{\ell=1,3}_{\rm kSZ}(k)$ is theoretically evaluated under the same WMAP9 cosmology as the ELEPHANT simulations. 

The fiducial values of $\sigma_v$ are chosen to be the averages of the best fitted $\sigma_v$ at different cut-off scales in the MCMC analysis of Sec.~\ref{subsubsec:MCMC_result}. The non-linear density power spectrum $P_{\delta\delta}$ is calculated by the Halofit operator of~\cite{Mead2015} embedded in the COLIBRI~\footnote{https://github.com/GabrieleParimbelli/COLIBRI} python package. The derivative $\partial X/\partial\theta_i$ is evaluated numerically by varying $\theta^0_i$, the fiducial value of $\theta_i$, with $\pm \lambda\theta^0_i$ ($\lambda=0.01$) and then 
\beq
\frac{\partial X}{\partial\theta_i} = \frac{X(\theta^0_i+\lambda\theta^0_i)-X(\theta^0_i-
\lambda\theta^0_i)}{2\lambda\theta^0_i}\,.
\eeq

\subsection{Monte Carlo analysis}
\label{subsec:MCMC}
\begin{figure*}
\includegraphics[width=1\columnwidth]{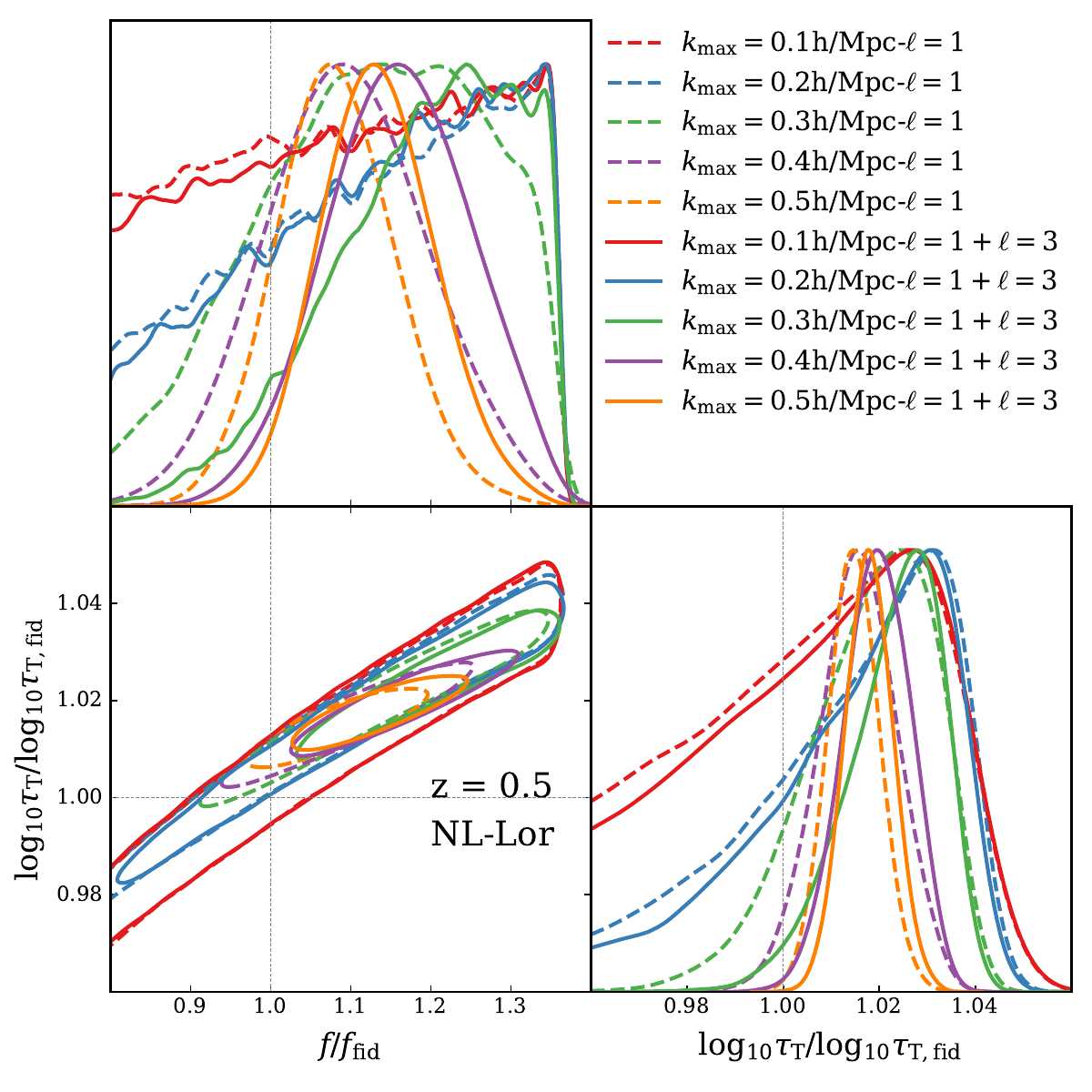}
\includegraphics[width=1\columnwidth]{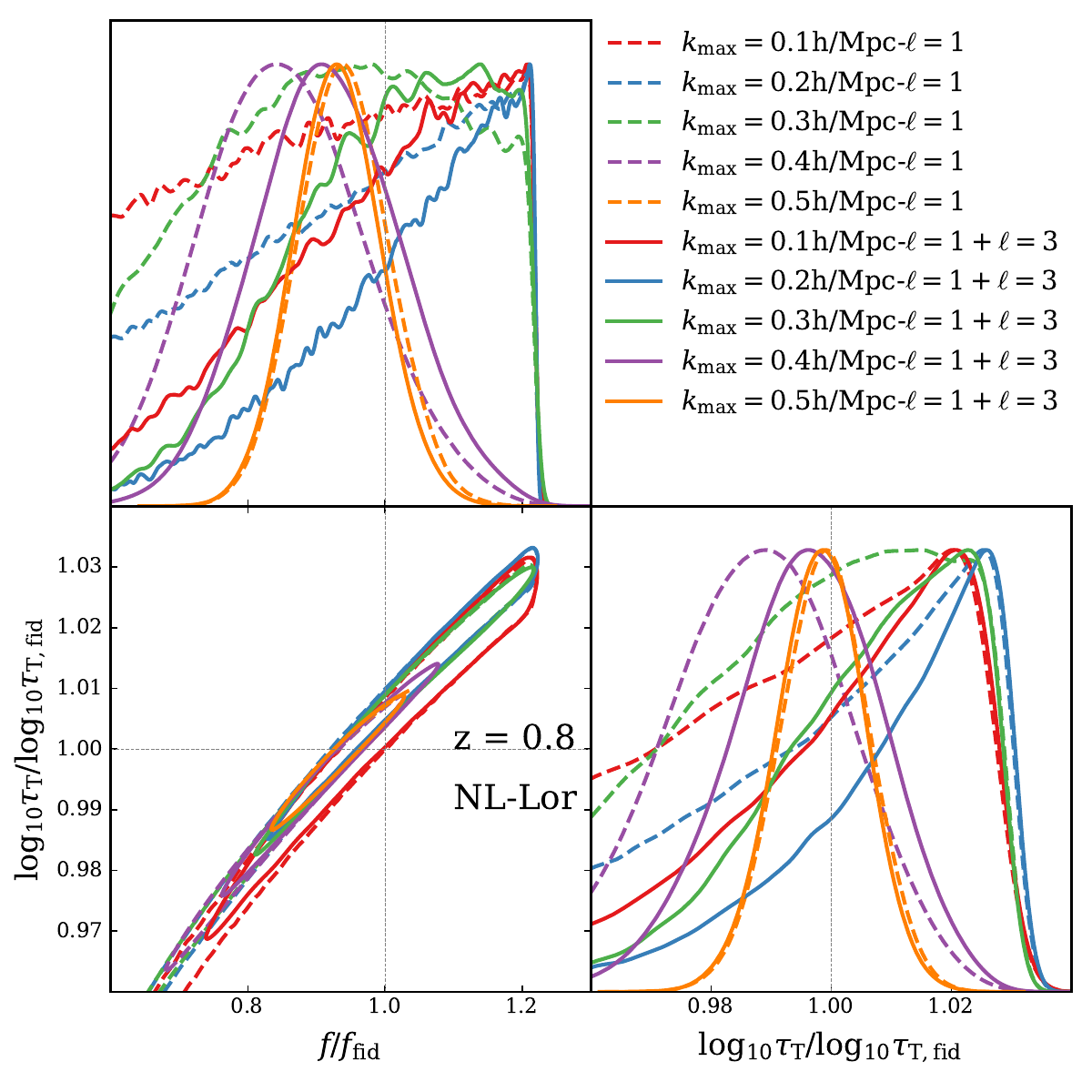}
\caption{{\it Left penal:} MCMC results for the NL-Lor model at $z=0.5$. {\it Right penal:} MCMC results for the NL-Lor model at $z=0.8$.{We adopt the non-linear $b(k)$ measured from simulations.} The solid $1-\sigma$ rang contours denote the MCMC analysis combining $P^{\ell=1}_{\rm kSZ}$ and $P^{\ell=3}_{\rm kSZ}$, and the dashed $1-\sigma$ range contours are for those with $P^{\ell=1}_{\rm kSZ}$ only. The marginalized 1-D PDFs of $f$ and $\log_{10}\tauT$ are shown on the top left and bottom right panels respectively. Different colors represent different cut-off scales $k_{\rm max}$ in the MCMC fitting.}
\label{fig:MCMC_fixedbk}
\end{figure*}

In Sec.~\ref{subsec:MCMC},  by \textsc{emcee}~\citep{emcee} we perform MCMC method to fit all four non-linear models presented in Sec.~\ref{subsec:nonlinear_models} against the measured $\Delta^{\ell=1,3}_{\rm kSZ}$ from simulations. 
In Sec.~\ref{subsec:fixed_bk_case}, the fitted parameters are $f$, $\log_{10} \tauT$ and $\sigma_v$, with uniform priors of $f \in [0,1]$, $\log_{10}\tauT \in [-7, -3]$ and $\sigma_v \in [0,1000]$ (Fig.~\ref{fig:MCMC_fixedbk}). {In Sec.~\ref{subsec:nonfixed_bk_case} we add $\{b_1,b_2\}$ as additional free parameters in the fitting, with either a wide flat prior being $b_1 \in [0,5]$ and $b_2\in[-5,15]$ (Fig.~\ref{fig:MCMC_nonfixedbk}) or a tight flat prior being $b_1/b_{1,\rm fid} \in [0.95,1.05]$ and $b_2/b_{2,\rm fid}\in[0.5,1.5]$ (Fig.~\ref{fig:MCMC_nonfixedbk_tight}).} 

The Likelihood function we choose is a common form, 
\begin{equation}
    \mathcal{L}_\ell(k_\mathrm{max}) \propto \exp{[-\chi^2_\ell(k_\mathrm{max})/2]} \label{eq:likelihood}
\end{equation}

\noindent where 
\begin{equation}
    \chi^2_\ell(k_{\mathrm{max}}) = \sum_{k=k_{\rm min}}^{k_{\mathrm{max}}} \frac{\left[\Delta_{\mathrm{kSZ, model}}^\ell(k) - \Delta_{\mathrm{kSZ, data}}^\ell(k) \right]^2}{{\sigma^{\ell}_{k}}^2}\,. \label{eq:chi2}
\end{equation}
\noindent $k_{\rm min}=0.035\hmpc$ and $k_{\mathrm{max}}$ denotes the max fitted $k$ mode between models and the mock data. As demonstrated in Appendix~\ref{app:covariance}, we ignore the off-diagonal elements of the covariance matrix and only use the diagonal variance ${\sigma^{\ell}_{k}}^2$ in evaluating the $\chi^2_\ell$. 


\section{results}
\label{sec:result}
\begin{figure*}
\includegraphics[width=\columnwidth]{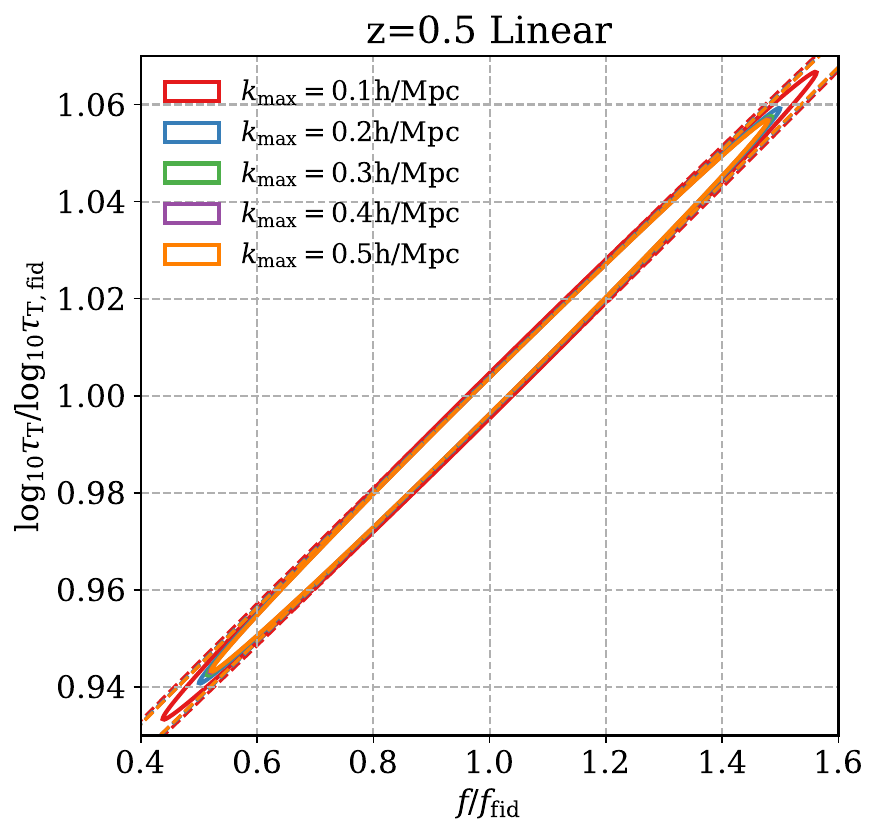}
\includegraphics[width=0.97\columnwidth]{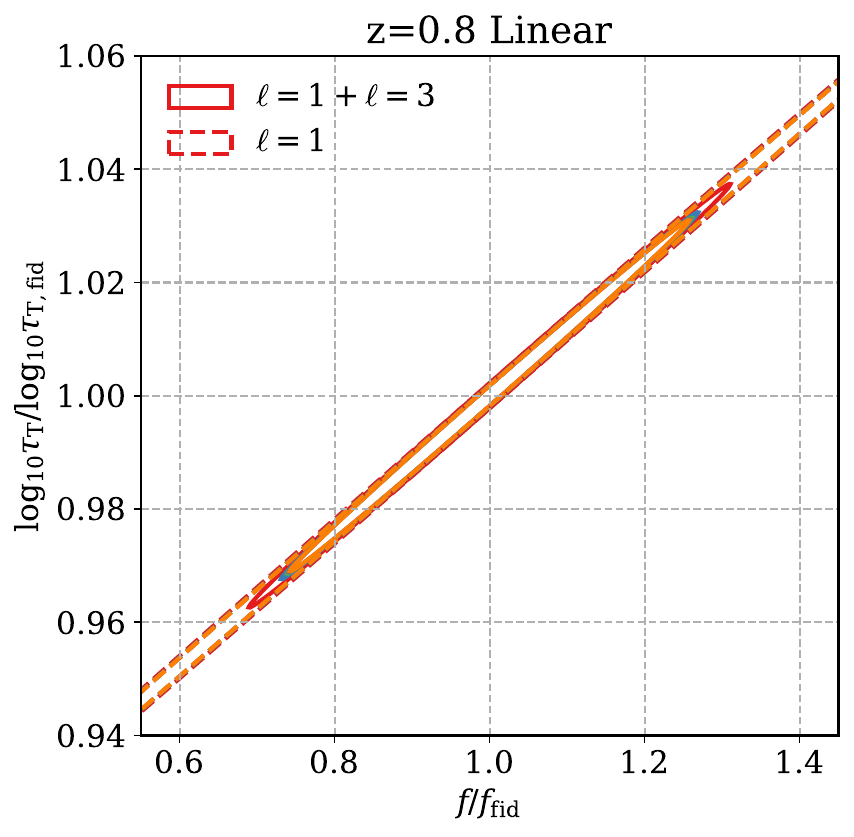}
\caption{Fisher matrix results for the linear model. {We adopt the non-linear $b(k)$ measured from simulations.} The solid $1-\sigma$ range contours denote the Fisher matrix analysis combining $P^{\ell=1}_{\rm kSZ}$ and $P^{\ell=3}_{\rm kSZ}$, and the dashed $1-\sigma$ rang contours are for those with $P^{\ell=1}_{\rm kSZ}$ only. Different colors represent different cut-off scales $k_{\rm max}$ in the Fisher matrix calculation.}
\label{fig:fisher_linear}
\end{figure*}
\begin{figure*}
\includegraphics[width=\columnwidth]{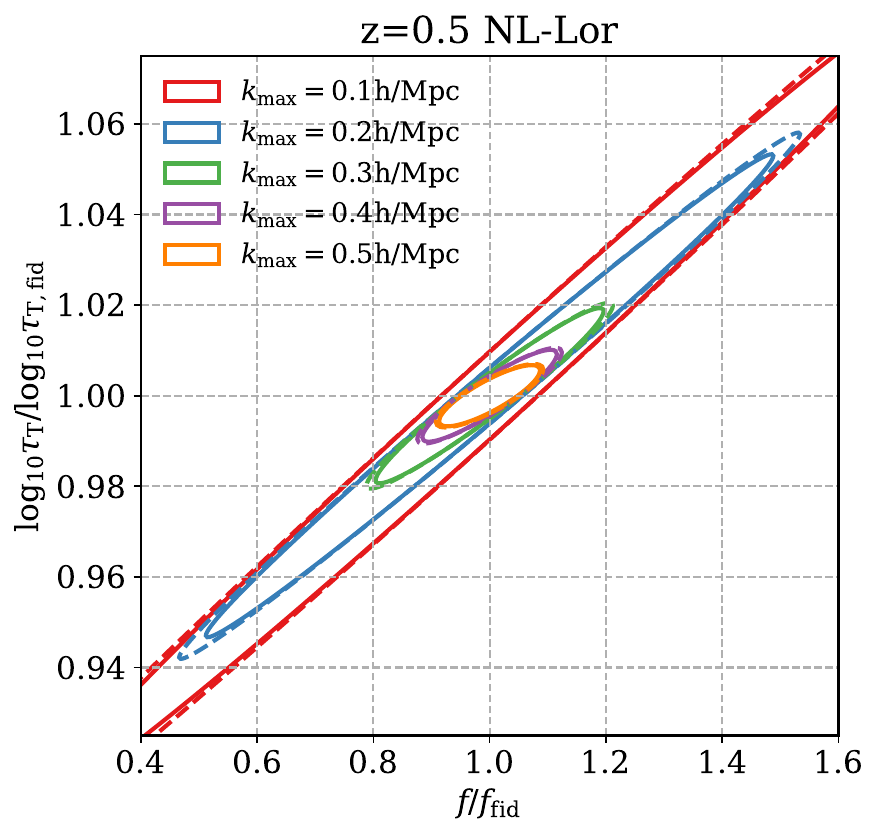}
\includegraphics[width=0.97\columnwidth]{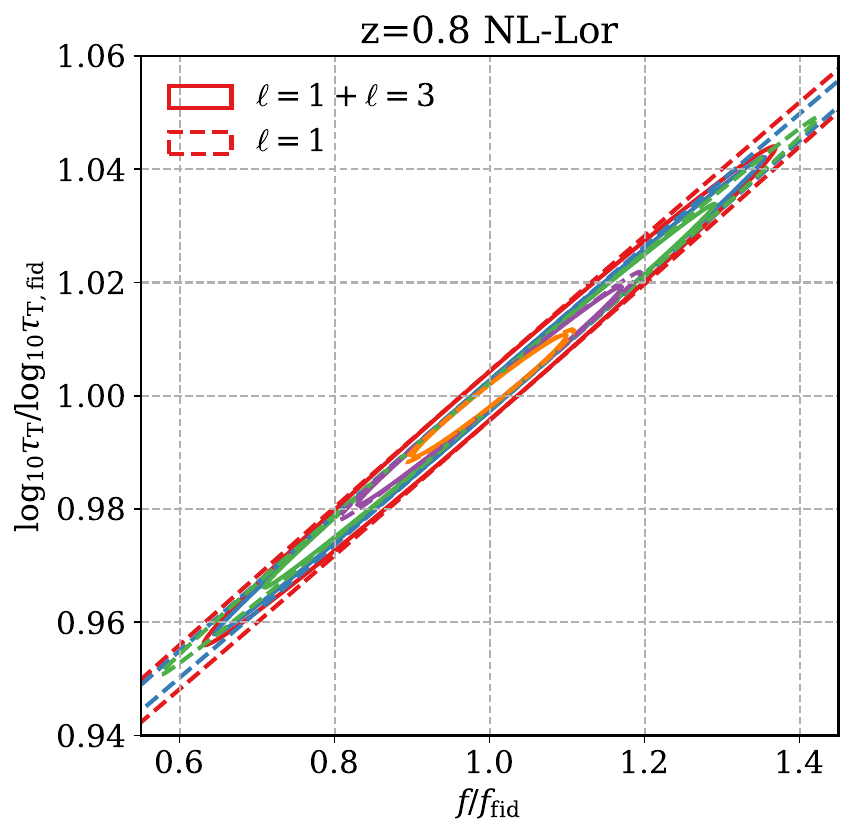}
\caption{Same as Fig.~\ref{fig:fisher_linear}, except for the NL-Lor model.}
\label{fig:fisher_nonlinear}
\end{figure*}

{In this section we present results of the Fisher matrix and MCMC analysis. We will first present results when we fix $b(k)$ as being measured from simulations. They clearly show the $\tauT-f$ degeneracy breaking in redshift space, in particular at non-linear scales. Then we show results when we include $b_1$ and $b_2$ in Eq.~(\ref{eq:NL_bk}) in the MCMC analysis. The latter results show that, unknown non-linear bias will weaken the level of $\tauT-f$ degeneracy breaking, unless we have tight priors on $b_1$ and $b_2$.}

\subsection{{The fixed \texorpdfstring{$b(k)$}{} case}}
\label{subsec:fixed_bk_case}
In this section we show the Fisher matrix and MCMC results when adopting the measured non-linear $b(k)$ from simulations.

\subsubsection{Fisher matrix results}
\label{subsubsec:fisher_result}

In Fig.~\ref{fig:fisher_linear}, we display the Fisher matrix constraints on $\log_{10}\tauT$ and $f$ with the linear model at different cut-off scales. 
By comparing the results using both $\Delta^{\ell=1,3}_{\rm kSZ}$ (`$\ell=1+\ell=3$' case) with those using only $\Delta^{\ell=1}_{\rm kSZ}$ (`$\ell=1$' case), we confirm that the RSD effect encoded in the measured $P^{\rm s}_{\rm kSZ}$ can break the $\tauT-f$ degeneracy and shrink the contour length along the degeneracy direction. Yet as we discussed in Sec.~\ref{subsec:linear_model}, in linear theory the degeneracy can be difficult to be further broken when we increase the cut-off scale in the analysis. This is justified in Fig.~\ref{fig:fisher_linear}, where the contours do not shrink notably with the cut-off scale $k_{\rm max}$ ranging from $0.1\hmpc$ to $0.5\hmpc$.

For a realistic estimation of the $\tauT-f$ contour at non-linear scales, we take a further step and utilize the non-linear models in Sec.~\ref{subsec:nonlinear_models} in the analysis. Four models give similar results. In Fig.~\ref{fig:fisher_nonlinear} we show the Fisher matrix contours from the `NL-Lor' model, as it is the relatively more accurate one within four variants in the MCMC analysis (as detailed in Appendix~\ref{app:accuracy}).

The further $\tauT-f$ degeneracy breaking due to the non-linear evolution of the density and velocity fields becomes manifest in Fig.~\ref{fig:fisher_nonlinear}. Although the width of the contour perpendicular to the degeneracy direction is enlarged due to the additional marginalized nuisance parameter $\sigma_v$, the degeneracy direction gradually rotates from $k_{\rm max}=0.1\hmpc$ to $k_{\rm max}=0.5\hmpc$, and the length of the contour considerably decreases. This is a clear evidence that, due to the different sensitivities of $P^{\rm s}_{\rm kSZ}$ with respect to $\tauT$ and $f$ generated by the non-linear evolution , we can robustly break the $\tauT-f$ degeneracy at non-linear scales in redshift space. 

By comparing the left and right panels of Fig.~\ref{fig:fisher_nonlinear}, we find a stronger degeneracy breaking on the left. This is probably due to the lower density bias of the $z=0.5$ mock than that of the $z=0.8$ one. The lower bias implies that the $bf\tauT$ term is less dominant in $P^{\rm s}_{\rm kSZ}$ so that the $\tauT-f$ degeneracy kept by this term is easier to be broken. This fact suggests that the $\tauT-f$ degeneracy breaking will be more efficient for galaxy samples with low density biases in a kSZ observation.

We also compare the `$\ell=1+\ell=3$' case with the `$\ell=1$' one in Fig.~\ref{fig:fisher_nonlinear}. The improvement of the parameter fitting by including $\Delta^{\ell=3}_{\rm kSZ}$ is larger for lower $k_{\rm max}$, and becomes smaller towards higher $k_{\rm max}$. It indicates that S/N of the kSZ measurement is dominated by $\Delta^{\ell=1}_{\rm kSZ}$ at small scales.


Finally, we notice that the forecasts for $f$ include unphysical results of $f>1$ in Figs.~\ref{fig:fisher_linear} and~\ref{fig:fisher_nonlinear}. This is because we practically (wrongly) apply an equivalent Gaussian prior with $0.5^2$ as the variance and $f_{\rm fid}(z)$ as the mean when we include the prior of $f$ in the Fisher matrix analysis. This unphysical phenomenon will be avoided in the MCMC results of Figs.~\ref{fig:MCMC_fixedbk}.

\subsubsection{MCMC fitting results}
\label{subsubsec:MCMC_result}

In order to confirm the Fisher matrix results in a more practical situation, we confront the non-linear theoretical models to the measured $\Delta^{\ell=1,3}_{\rm kSZ}$ of the mock galaxies using the MCMC technique in this section. We simultaneously fit $\{\log_{10}\tauT,f,\sigma_v\}$ in the analysis and the fitting results of $\{\log_{10}\tauT,f\}$ are shown in Figs.~\ref{fig:MCMC_fixedbk}.

As we can see from the lower left sub-panels of Figs.~\ref{fig:MCMC_fixedbk}, the fitted contours of MCMC results gradually turn their degeneracy directions when the adopted $k_{\rm max}$ becomes higher, along with the shrinking of the contour sizes. This is consistent with the Fisher matrix forecasts. It further consolidates our statement that the non-linear structure evolution in redshift space break the $\tauT-f$ degeneracy in a robust way.

We also compare MCMC results using both $\Delta^{\ell=1,3}_{\rm kSZ}$ with those using only $\Delta^{\ell=1}_{\rm kSZ}$. Similar to the Fisher matrix analysis, the additional information of $\Delta^{\ell=3}_{\rm kSZ}$ helps tighten the $\tauT-f$ constraint at small $k_{\rm max}$ cases, while the improvement diminishes when $k_{\rm max}$ is large.

There is a clear difference between MCMC and Fisher matrix results. In Figs.~\ref{fig:MCMC_fixedbk}, there are evident cuts at $f=1$ on the $\log_{10}\tauT-f$ contours. This is due to the correct implementation of the flat $f\in[0,1]$ prior in the MCMC analysis.

Furthermore, a unique advantage of the MCMC analysis is that besides the precision forecast, it can test the model accuracy as well. As shown in Figs.~\ref{fig:MCMC_fixedbk}, the NL-Lor model is accurate within roughly $1-2\sigma$ level in predicting $\log_{10}\tauT$ and $f$ even at non-linear scales of $k>0.3\hmpc$. Considering that our models are simple and phenomenological, this level of accuracy is encouraging and we expect that further refinements on the current models will improve their accuracy. The detailed accuracy comparison of different non-linear models is shown in Appendix~\ref{app:accuracy}.

\subsection{{The non-fixed \texorpdfstring{$b(k)$}{} case}}
\label{subsec:nonfixed_bk_case}
\begin{figure*}
\includegraphics[width=1\columnwidth]{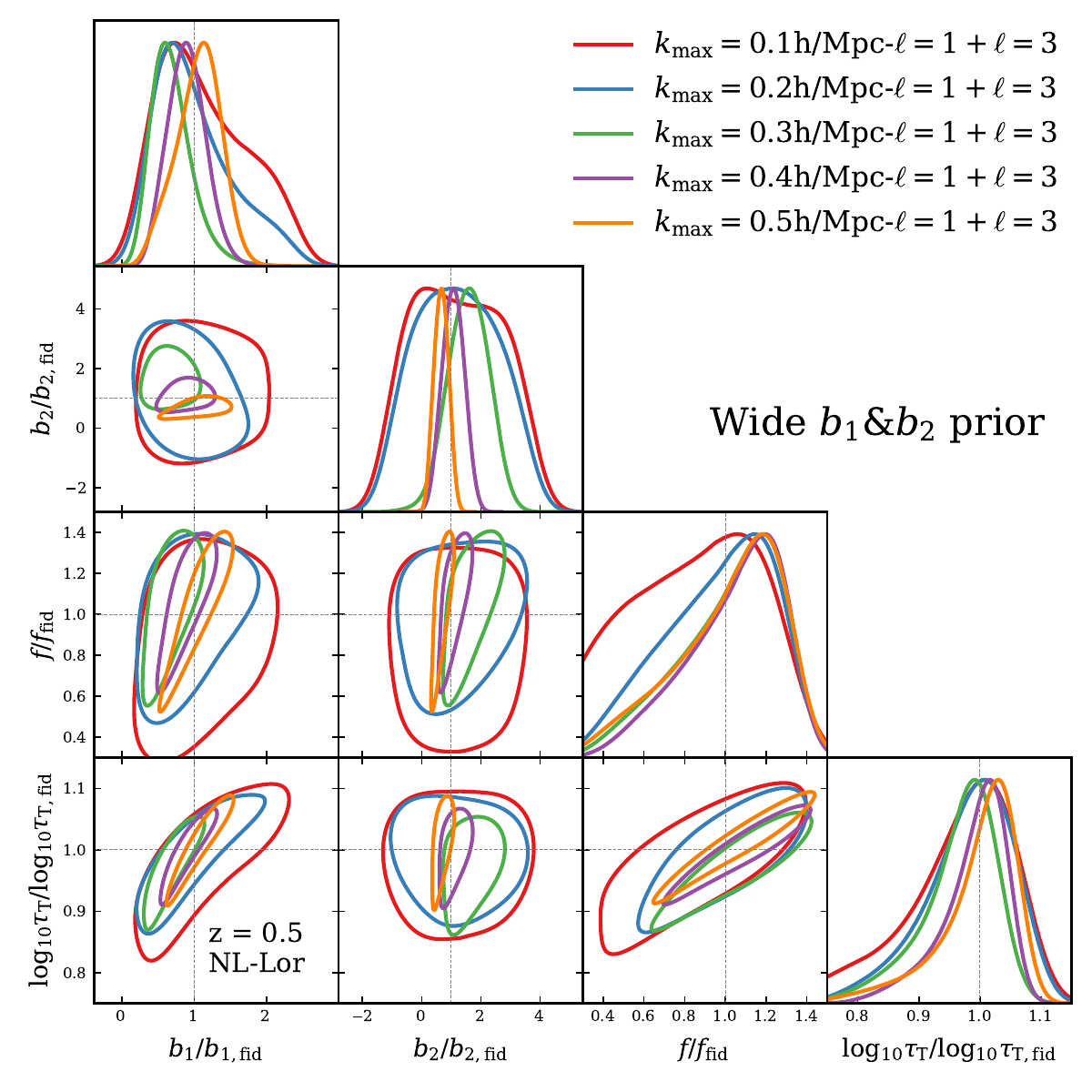}
\includegraphics[width=1\columnwidth]{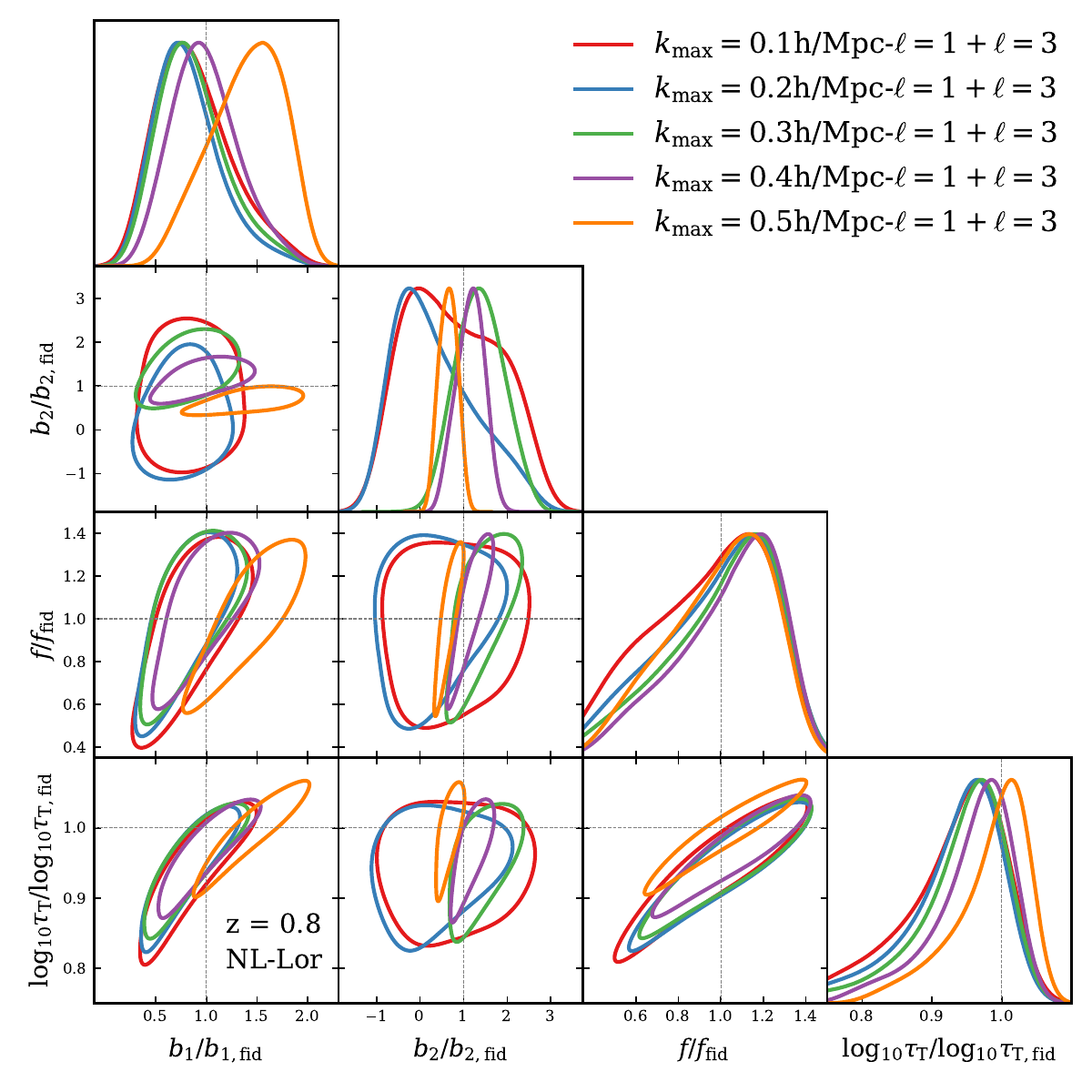}
\caption{{\it Left penal:} {MCMC results for the NL-Lor model at $z=0.5$ when we include $b_1$ and $b_2$ in Eq.~(\ref{eq:NL_bk}) as free parameters. {\it Right penal:} the same as the left panel, except at $z=0.8$. Different colors represent different cut-off scales $k_{\rm max}$ in the MCMC fitting. Here set a wide flat prior of $b_1\in[0,5]$ and $b_2\in[-5,15]$.}}
\label{fig:MCMC_nonfixedbk}
\end{figure*}
\begin{figure*}
\includegraphics[width=1\columnwidth]{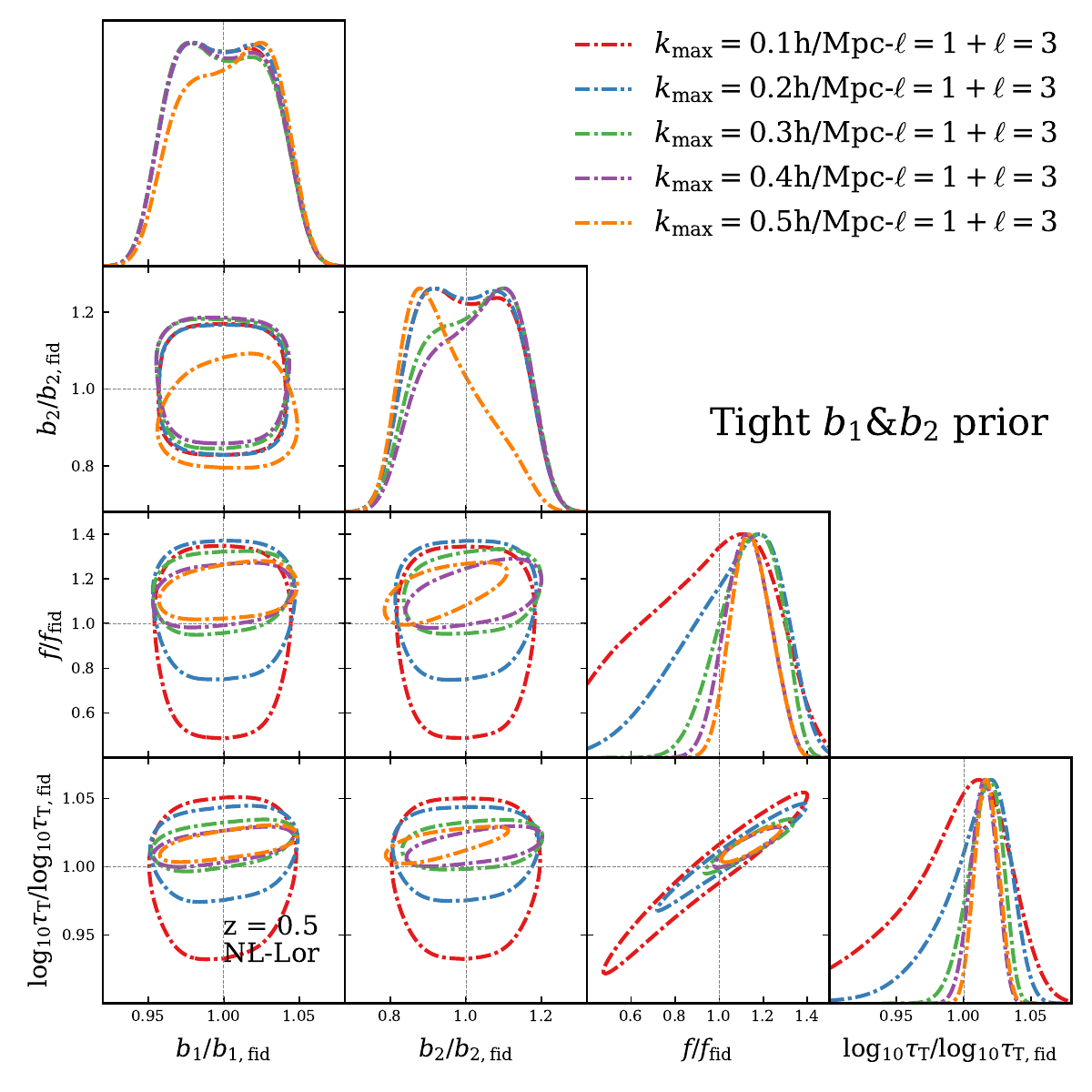}
\includegraphics[width=1\columnwidth]{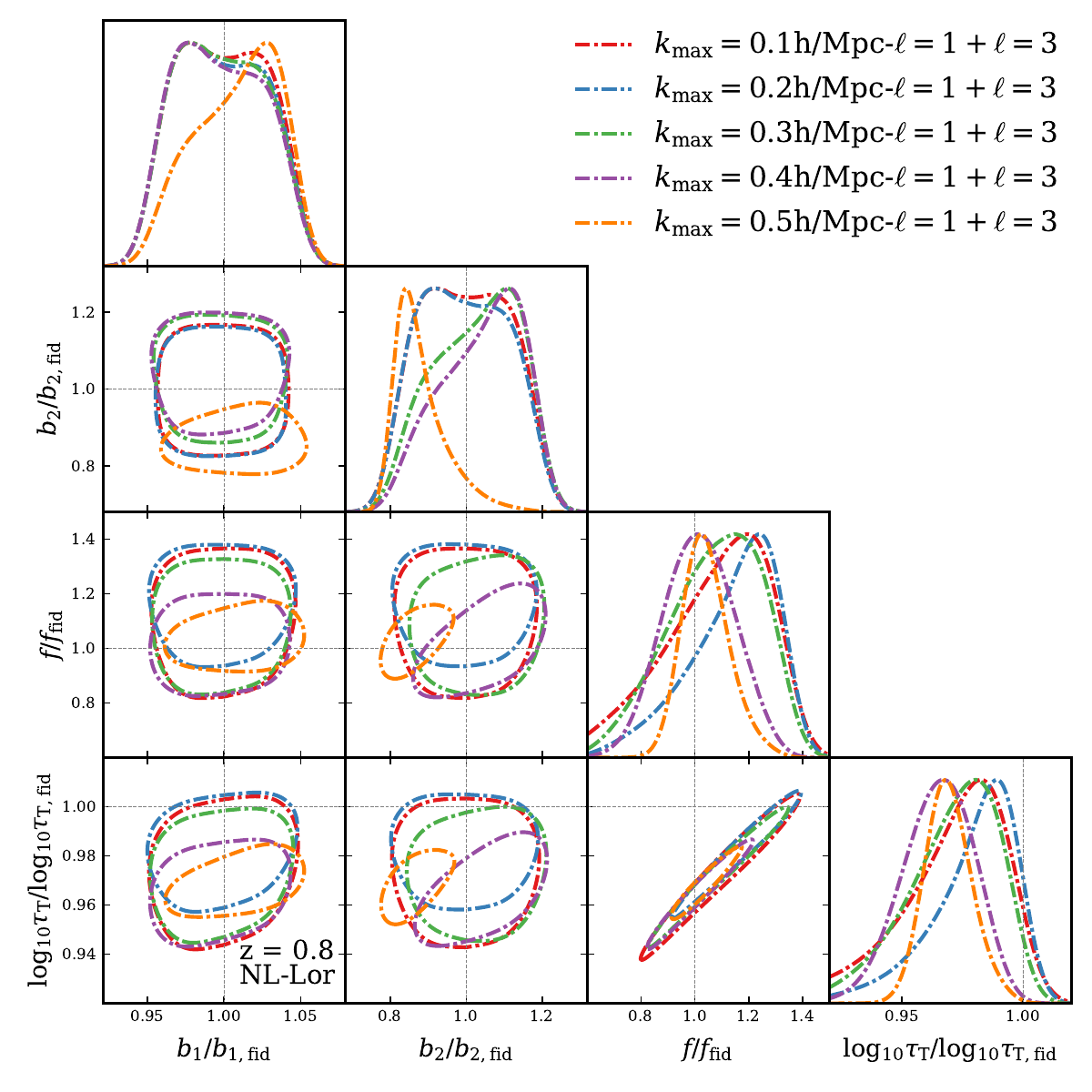}
\caption{{The same as Fig.~\ref{fig:MCMC_nonfixedbk}, except that we set a tight flat prior being $b_1/b_{1,\rm fid} \in [0.95,1.05]$ and $b_2/b_{2,\rm fid}\in[0.5,1.5]$ in the MCMC fitting.}}
\label{fig:MCMC_nonfixedbk_tight}
\end{figure*}

{In a real kSZ data analysis, the unknown non-linear density bias brings unavoidable complications. As shown in Eq.~(\ref{eq:P_kSZ_lin_l1}), the linear density bias $b_1$ is degenerated with $f$ and $\tauT$ at linear scales, and this degeneracy still exists on non-linear scales in Eqs.~(\ref{eq:P_kSZ_Gau}) and~(\ref{eq:P_kSZ_Lor}). In this sense, we show in this section MCMC results when we include $b_1$ and $b_2$ in Eq.~(\ref{eq:NL_bk}) as free parameters in the fitting process.}

{In Fig.~\ref{fig:MCMC_nonfixedbk} we show the posterior contours of $[b_1,b_2,f,\log_{10}\tauT]$. We can still see the mild $\tauTf$ degeneracy breaking at $k\le 0.3 \hmpc$, yet it is stopped at $k_{\rm max}>0.3\hmpc$. Only contours of $b_1$ and $b_2$ continue to shrink at $k_{\rm max}>0.3\hmpc$. This can be due to the degeneracy between $f$, $\tauT$ and $b(k)$ and it illustrates the importance of a well understanding of $b(k)$ before we intend to break the $\tauTf$ degeneracy at non-linear scales.  In Appendix~\ref{app:MCMC_nonfixb1}, we present the results when we adopt a linear bias model and only include $b_1$ as an additional parameter. There we yield similar conclusions, except that the wrong (linear) bias model at non-linear scales further induces strong systematic errors of the fitted $f$ and $\log_{10}\tauT$ at $z=0.8$.}

{As a further illustration of the importance of $b(k)$, we put a tight prior on $b_1$ and $b_2$, and redo the MCMC fitting. The results are shown in Fig.~\ref{fig:MCMC_nonfixedbk_tight}. As expected, a tight prior, namely a well understanding of $b(k)$, restores the $\tauTf$ degeneracy breaking at non-linear scales, although the inaccurate $b(k)$ model induces severe systematic error on $f$ and $\tauT$ at $z=0.8$.}

{Therefore, we see that a good knowledge of the non-linear bias term is essential for the cosmological application of the kSZ effect. This information will require external data such as the angular galaxy clustering and weak lensing observation (e.g.~\citealt{ShaoZW2023}).}

\section{Conclusion and discussion}
\label{sec:discussion}

In a late-time kSZ effect observation, we apply the kSZ tomography technique in redshift space where the velocity field leaves its own distinct imprint on the galaxies' redshift space positions. The extra information of the linear growth rate $f$ can thus be provided by the RSD effect encoded in the observed kSZ effect itself, allowing for the robust breaking of the $\tauT-f$ degeneracy. We adopt the Fisher matrix and MCMC techniques to validate this idea in this work. Both methods yield positive and consistent results, finding that the level of this degeneracy breaking is further enhanced on non-linear scales due to the non-linear evolution of the density and velocity fields. This result highlights the importance of the redshift space analysis of the pairwise kSZ effect in a kSZ observation, in particular on non-linear scales.

{Yet, a good knowledge of the non-linear density bias of galaxies is needed before the $\tauTf$ degeneracy can be robustly broken. This will require external data such as such as the angular galaxy clustering and weak lensing observation (e.g.~\citealt{ShaoZW2023}). In this sense, the synergy of the kSZ effect with other cosmological probes are essential for its cosmological application, where the non-linear bias information can be provided externally. Meanwhile, an accurate non-linear bias model is needed to guarantee unbiased determinations of $f$ and $\tauT$ values.}

In order to fully explore its cosmological information, an accurate non-linear model of the kSZ effect in redshift space is also needed. We develop a phenomenological model of the density-weighted pairwise kSZ power spectrum in this work. For a DESI+CMB-S4-like survey combination, its fitted values of the linear growth rate and the optical depth are accurate within $1-2\sigma$ ranges of the fiducial ones, even down to small scales around $k=0.5\hmpc$.

Alternative models of the pairwise kSZ effect exist in the literature. For example, in \cite{Sugiyama2016}, authors applied $P^{\rm s}(k)$ calculated by the Lagrangian perturbation theory at zeroth-(Zel'dovich approximation) and one-loop order to Eq.~(\ref{eq:P_kSZ_der}) and derived $P^{\rm s}_{\rm pv}$. The model worked well for dark matter and can be used for halos by incorporating scale dependent density and velocity bias models. In \cite{Okumura2014}, the authors studied  the density-weighted velocity statistics in redshift space by the distribution function approach (DFA), in which one-loop SPT and a free parameter describing the nonlinear velocity dispersion were adopted. This model has been successfully utilized to describe the observed pairwise kSZ correlation function from BOSS+ACT~\citep{DeBernardis17}. Furthermore, the deep learning technique has been proposed to reconstruct the velocity field from the kSZ effect, avoiding an independent estimation of the optical depth~\citep{WangYY2021}.

As a summary, we expect that the $\tauT-f$ degeneracy can be robustly broken by carefully studying the pairwise kSZ effect in  redshift space and on non-linear scales, {if we have a good knowledge of the non-linear galaxy density bias. Therefore, extra information of the cosmological structure formation can be extracted from the non-linear kSZ observation, and the kSZ effect will serve as a good supplementary probe of cosmology in the future.}

\section*{Acknowledgements}
We thank Zhuoyang Li, Pengjie Zhang and Kwan Chuen Chan for useful discussions. {We thank the referee Naonori Sugiyama for bringing up valuable comments and heavily improve the paper quality.} We appreciate Baojiu Li and Minji Oh for providing the simulation and catalog data. YZ acknowledges the supports from the National Natural Science Foundation of China (NFSC) through grant 12203107, the Guangdong Basic and Applied Basic Research Foundation with No.2019A1515111098, and the science research grants from the China Manned Space Project with NO.CMS-CSST-2021-A02. Numerical calculations were performed by using a high performance computing cluster of the Department of Astronomy, Sun Yat-Sen University and a high performance computing cluster in the Korea Astronomy and Space Science Institute.

\section*{Data Availability}
The data underlying this article are available on request to the corresponding author.


\bibliographystyle{mnras}
\bibliography{mybib} 

\appendix

\section{Proof of the \texorpdfstring{$\tauT-\MakeLowercase{f}$}{} degeneracy on linear scales}
\label{app:degeneracy_proof}
Required by the equivalence principle, the galaxy velocity field locally coincides with that of its ambient dark matter, and statistically converges to the latter on linear scales~\citep{Desjacques10,Desjacques18biasreview,Zheng14b,Junde18vb}. Thus on linear scales, we can replace the galaxy velocity field $\bfv_g$ with the dark matter velocity field $\bfv$, and the measured $\delta T_{\rm kSZ}$ field can be linearized as
\bea
\label{eq:Tksz_field}
\sum_{i=1}^{N}\delta T_{{\rm kSZ},i} &=&-\frac{T_0\tauT}{c}\sum_{i=1}^{N}\bfv_i\cdot\hat{n}_i \no\\
&\propto&-\frac{T_0\tauT}{c}[1+\delta_g(\bfx)]v_{\hat{n}}(\bfx)\approx-\frac{T_0\tauT}{c}v_{\hat{n}}(\bfx)\,.\no
\eea
Here $v_{\hat{n}}\equiv\bfv\cdot\hat{n}$ is the LOS velocity field. The linearized continuity equation
\beq
\label{eq:continuity}
\frac{\partial\delta(\bfx)}{\partial t}+\frac{1}{a}\nabla\cdot\bfv(\bfx)=0
\eeq
can be reformed in Fourier space as
\beq
\label{eq:continuity_fourier}
afH\delta(\bfk)+i\bfk\cdot \bfv(\bfk) =0\,,
\eeq
where $f\equiv d\ln D/d\ln a$ is the linear growth rate and $D$ is the linear growth factor. We solve Eq.~(\ref{eq:continuity_fourier}) for
\beq
v_{\hat{n}}(k,\mu)= i\mu\frac{afH}{k}\delta(k)\,.
\eeq
$\mu\equiv\cos(\hat{\bfk}\cdot\hat{n})$ denotes the cosine of the angle between $\bfk$ and the LOS, and then the measured $\delta T_{\rm kSZ}$ field is
\beq
\label{eq:Tksz_field2}
\sum_{i=1}^N T_{{\rm kSZ},i} \propto -i\mu\frac{T_0aH}{ck}f\tauT\delta(k)\propto f\tauT\,.
\eeq
In this sence we prove the $\tauTf$ degeneracy on linear scales in real space.

\section{Proof of Eq.~(7)}
\label{app:eq_ksz_der_proof}

We treat tracers as discretized objects. Their number density in redshift space is
\beq
n(\bfs) = \sum_{i=1}^{N} \delta_{\rm D}(\bfs-\bfs_i)\,,
\eeq
where $N$ is the galaxy number and $\delta_{\rm D}$ is the Dirac Delta function. The density fluctuation in Fourier space is then
\beq
\delta^{\rm s}(\bfk) = \sum_{i=1}^{N}e^{-i\bfk\cdot\bfs_i}\,.
\eeq
Omitting the normalization prefactor, the density power spectrum is defined as
\beq
\delta_{\rm D}(\bfk+\bfk')P^{\rm s}(\bfk) \equiv \left<\delta^{\rm s}(\bfk)\delta^{\rm s}(\bfk')\right>\,.\no
\eeq
Then we have 
\bea
P^{\rm s}(\bfk) &=& \left<\sum_{i,j=1}^{N} e^{-i\bfk\cdot(\bfs_i-\bfs_j)}\right>\no \\
&=&\left<\sum_{i,j=1}^{N} e^{-i\bfk\cdot(\bfx_i-\bfx_j)-ik\mu\frac{\bfv_i\cdot\hat{n}_i-\bfv_j\cdot\hat{n}_j}{aH}}\right>\,.
\eea
Considering $\bfv\propto f$, we have
\bea
\frac{\partial}{\partial f} P^{\rm s}(\bfk) &=& \frac{-ik\mu}{afH}\left<\sum_{i,j=1}^{N}(\bfv_i\cdot\hat{n}_i-\bfv_j\cdot\hat{n}_j) e^{-i\bfk\cdot(\bfs_i-\bfs_j)}\right>\no \\
&\equiv&\frac{-ik\mu}{afH}P^{\rm s}_{\rm pv}\,.
\eea
Then we prove Eq.~(\ref{eq:P_kSZ_der}):
\beq
P^{\rm s}_{\rm pv}(\bfk) = \left(i\frac{aHf}{\bfk \cdot \hat{n}}\right) \frac{\partial}{\partial f} P^{\rm s}(\bfk) \,.
\eeq

We notice that the relation $\bfv\propto f$ is only {an approximation} on non-linear scales. Considering the good performance of the non-linear models derived from Eq.~(\ref{eq:P_kSZ_der}) in this work, we regard $\bfv\propto f$ as a good approximation here and leave further refinements on it to future works.

The detailed derivation of this relationship, including its extensions to higher order moments of density-weighted velocity field, can be found in~\cite{Sugiyama2016,Sugiyama2017}.

\section{Covariance matrix}
\label{app:covariance}
\begin{figure}
\begin{center}
\includegraphics[width=\columnwidth]{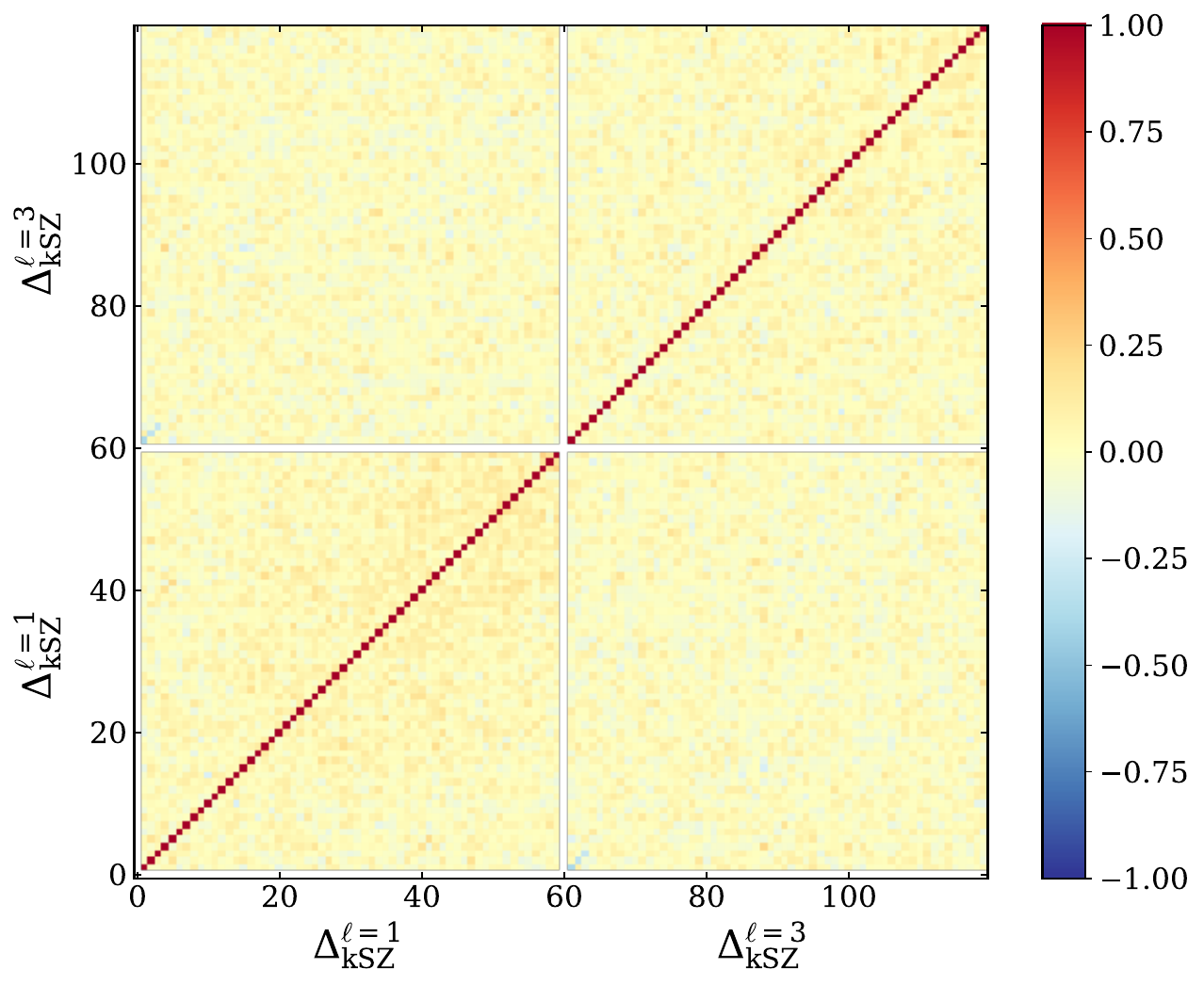}
\end{center}
\caption{Measured correlation coefficients of $\Delta^{\ell=1,3}_{\rm kSZ}$ from 100 simulations. The $k$ range is $k\in[0.03-0.6]\hmpc$, with a bin width of $\Delta k=0.01\hmpc$.}
\label{fig:cov}
\end{figure}
\begin{figure}
\begin{center}
\includegraphics[width=\columnwidth]{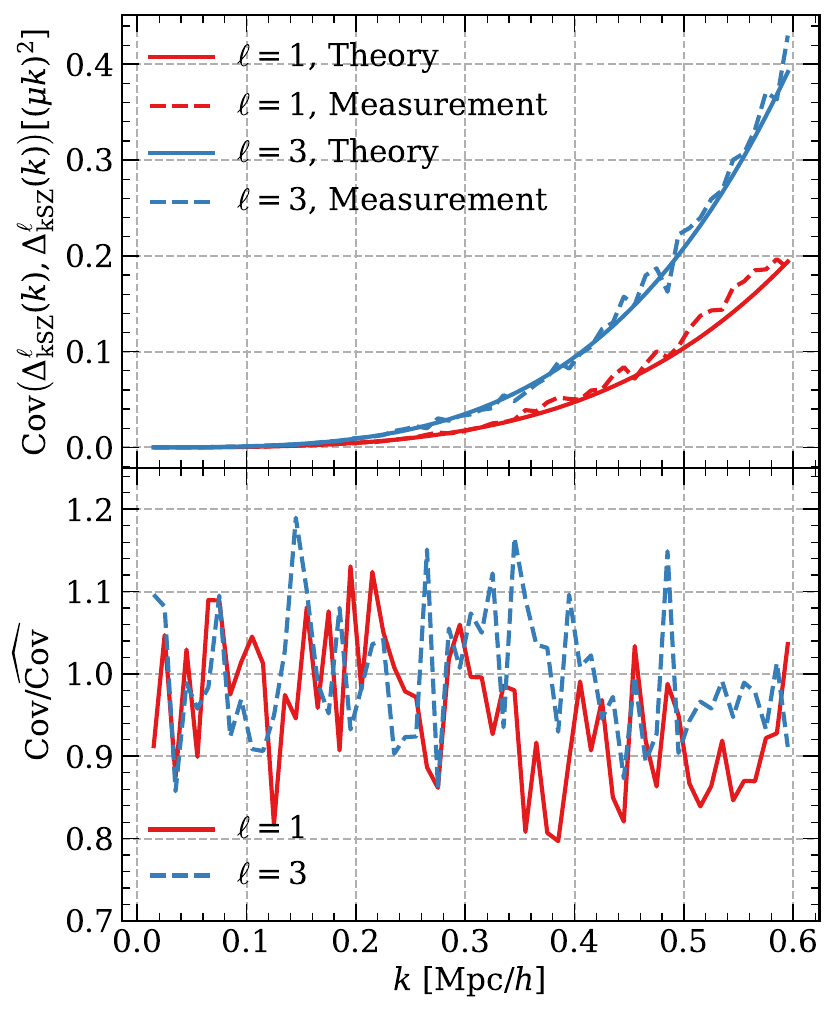}
\end{center}
\caption{{\it Top panel:} the diagonal elements of the $\Delta^{\ell=1,3}_{\rm kSZ}$ covariance matrix theoretically calculated by Eq.~(\ref{eq:cov}) and measured from simulations by Eq.~(\ref{eq:cal_cov}). {\it Bottom panel:} the ratio between the theory and the simulation measurement.}
\label{fig:cov_ratio}
\end{figure}
\begin{table}
\begin{center}
\begin{tabular}{r|c|c|c|cccccc}\hline\hline
redshift & $\log_{10}M$ range & $N_h/10^5$ &$n_h$ &$b_1$\\\hline
$z=0.9$\ \  &  $13.0-13.5$ & 11 & 1.6 & 2.4 \\\hline\hline
\end{tabular}
\end{center}
\caption{The properties of the test halo sample. The logarithmic mass unit is $M_\odot/h$ and the halo number density $n_h$ has unit of $10^{-4}(\mpch)^{-3}$.  $N_h$ is the total halo number in a halo mass bin. $b_1$ is the linear density bias. }
\label{tab:halo}

\end{table}

Following~\cite{Sugiyama2017}, we ignore the off-diagonal elements of the matrix, and the covariance matrix between $\Delta_{\rm kSZ}^{\ell_1}$ and $\Delta_{\rm kSZ}^{\ell_2}$ can be calculated by
\bea
\label{eq:cov}
&&{\rm Cov}(\Delta_{\rm kSZ}^{\ell_1}(k), \Delta_{\rm kSZ}^{\ell_2}(k)) \no\\
&=&\left(\frac{k^3}{2\pi^2}\right)^2\frac{1}{N_{\rm mode}(k)}\frac{2(2\ell_1+1)(2\ell_2+1)}{4\pi} \no\\
&\times&\int d\varphi \int d\mu {\cal L}_{\ell_1}(\mu){\cal L}_{\ell_2}(\mu)\left(\frac{T_0\tauT}{c}\right)^2\left[2\left(P^{\rm s}_{(1)(0)}(\bfk)\right)^2\right.\no\\
&-&\left.2\left(P^{\rm s}_{(1)(1)}(\bfk)+(1+R_{\rm N}^2)\frac{\sigma^{(2)}_{\rm v}}{\bar{n}}\right)\left(P^{\rm s}(\bfk)+\frac{1}{\bar{n}}\right)\right]\,.
\eea
Here $N_{\rm mode}(k)$ is the number of $k$ modes in a given $k$ bin. ${\cal L}_{\ell}(\mu)$ is the Legendre polynomial. $\bar{n}$ is the number density of mock galaxies. $\sigma_{\rm v}^{(2)}$ is the measured (density-weighted) velocity dispersion of mock galaxies from simulations. In redshift space, $P^{\rm s}_{(1)(0)}$ is the momentum-density cross-power spectrum, $P^{\rm s}_{(1)(1)}$ is the momentum-momentum auto-power spectrum, and $P^{\rm s}$ is the density-density auto-power spectrum. All three power spectra are also calculated from simulations.

In particular, we define $R_{\rm N}$, the inverse S/N, as
\beq
R^2_{\rm N}(\theta_{\rm c}) = \frac{\sigma^2_{\rm N}(\theta_{\rm c})}{\sigma_{\rm kSZ}^2(\theta_{\rm c})}\,,
\eeq
with the signal
\beq
\sigma_{\rm kSZ}(\theta_{\rm c}) = \left(\frac{T_0\tauT(\theta_{\rm c})}{c}\right)\sigma_{\rm v}\,,\,\,\,\,\, \sigma_{\rm v} = \sqrt{\sigma_{\rm v}^{(2)}}\,,
\eeq
and the noise
\beq
\sigma_{\rm N}^2(\theta_{\rm c}) = \sum^{\ell_{\rm max}}_\ell\frac{2\ell+1}{4\pi}\langle C^{\rm obs}_\ell\rangle U(\ell\theta_{\rm c}) U(\ell\theta_{\rm c}) \,,
\eeq
in which $U(\ell\theta_{\rm c})$ is the Fourier transform of the AP filter~\citep{David2016}. The ensemble average of the observed CMB angular power spectrum is $\langle C^{\rm obs}_\ell\rangle = B^2_\ell C^{\rm the}_\ell + N_\ell$, with a Gaussian beam function $B_\ell$,  the theoretical prediction of the CMB power spectrum $C^{\rm the}_\ell$ including the CMB lensing and the detector noise $N_\ell$. We choose the $\theta_{\rm c}$ that minimizes $R_{\rm N}$ and calculate the corresponding $\tauT$ and ${\rm Cov}(\Delta_{\rm kSZ}^{\ell_1}(k), \Delta_{\rm kSZ}^{\ell_2}(k))$.

In Eq.~(\ref{eq:cov}), only the Gaussian part of the covariance matrix are considered. In principle, non-Gaussian terms will become more important at smaller scales, enhance the diagonal components and generate off-diagonal components. Since we work on non-linear scales with $k$ up to $0.5\hmpc$, it is necessary to quantify the accuracy of this Gaussian approximation. For doing this, we directly measure the covariance of $\Delta^{\ell=1,3}_{\rm kSZ}$ from a suite of 100 N-body simulations, and compare it with the theoretical calculation from Eq.~(\ref{eq:cov}). {Besides this,} we {also notice} that the robustness of Eq.~(\ref{eq:cov}) has also been demonstrated in the Appendix A of~\citep{Sugiyama18} by using Jackknife and Bootstrap re-sampling methods on real observational data.

The adopted 100 N-body simulations~\citep{Zheng16a} are run by \texttt{Gadget-2}~\citep{Springel05} under a Planck2015 $\Lambda$CDM cosmology~\citep{PLANK2015}. It has a large box size of $(1.89{\rm Gpc}/h)^3$ and $1024^3$ particles. We generate halo catalogs by the \texttt{ROCKSTAR} halo finder~\citep{Rockstar}, and select the halo sample with mass $M_h\in [10^{13},10^{13.5}]M_{\odot}/h$ at $z=0.9$ as the test sample~\citep{ZhengY2019}. The properties of this test sample is listed in Table~\ref{tab:halo}.

The measured kSZ temperature fluctuation of the $i$th halo, including all noise components, can be expressed as
\beq
\label{eq:mock_TkSZ}
\delta \hat{T}_{{\rm kSZ},i} = -\left(\frac{T_0\tauT}{c}\right)\bfv_i\cdot \hat{n}_i+\delta T_{{\rm N},i}\,.
\eeq
Here the noise term $\delta T_{{\rm N}}$ is independent of the signal, and assumed to be an uncorrelated, Gaussian field with zero mean and a standard deviation $\sigma_{\rm N}$.

We set $\tauT=6\times10^{-5}$ and $\sigma_{\rm N} = 2\mu K$. For each halo, we randomly select a $\delta T_{{\rm N},i}$ from the Gaussian distribution of $\delta T_{\rm N}$ and generate $\delta \hat{T}_{{\rm kSZ},i}$ by Eq.~(\ref{eq:mock_TkSZ}). Along each LOS, we then calculate $\Delta^{\ell=1,3}_{\rm kSZ}$ in each simulation and evaluate the covariance $\widehat{\rm Cov}(\Delta^{\ell_1}_{\rm kSZ},\Delta^{\ell_2}_{\rm kSZ})$ by
\bea
\label{eq:cal_cov}
&&\widehat{\rm Cov}(\Delta^{\ell_1}_{\rm kSZ}(k_1),\Delta^{\ell_2}_{\rm kSZ}(k_2)) \no\\
&=& \frac{1}{N-1} \sum^{N}_{i=1}\left[\left(\widehat{\Delta^{\ell_1}_{{\rm kSZ},i}}(k_1)-\overline{\Delta^{\ell_1}_{\rm kSZ}}(k_2)\right)\right.\no\\
&&\times\left.\left(\widehat{\Delta^{\ell_2}_{{\rm kSZ},i}}(k_1)-\overline{\Delta^{\ell_2}_{\rm kSZ}}(k_2)\right)\right]\,,
\eea
in which $N=100$ and $\overline{\Delta^{\ell}_{\rm kSZ}}$ is the mean of measured power spectra from 100 simulations. The measured covariance matrices along 3 LOS's are further averaged to reduce its uncertainty.

Fig.~\ref{fig:cov} shows the measured correlation coefficients,
\beq
r^{\ell_1\ell_2}_{k_1k_2} = \frac{\widehat{\rm Cov}\left(\Delta^{\ell_1}_{\rm kSZ}(k_1),\Delta^{\ell_2}_{\rm kSZ}(k_2)\right)}{\sigma_{\Delta^{\ell_1}_{\rm kSZ}}(k_1)\sigma_{\Delta^{\ell_2}_{\rm kSZ}}(k_2)}\,,
\eeq
where $\sigma^2_{\Delta^{\ell}_{\rm kSZ}}(k)$ is the measured variance of $\Delta^{\ell}_{\rm kSZ}(k)$. Although the measurements are noisy due to a limited number of simulations, we do not observe significant off-diagonal elements, in particular for covariance between $\Delta^{\ell=1}_{\rm kSZ}$ and $\Delta^{\ell=3}_{\rm kSZ}$. We check that all the off-diagonal correlation coefficients are $<0.3$. Therefore ignoring off-diagonal elements of the covariance in the analysis is not expected to significantly bias our results in this work.

In Fig.~\ref{fig:cov_ratio}, we compare the diagonal elements of the theoretical covariance of $\Delta^{\ell=1,3}_{\rm kSZ}$ ($\rm Cov$) with those from the measurement ($\widehat{\rm Cov}$). We are using these diagonal elements in our Fisher matrix and Monte Carlo analysis. For $\Delta^{\ell=1}_{\rm kSZ}$, $\rm Cov/\widehat{Cov}$ gradually drops from 1 at $k<0.3\hmpc$ to 0.9 at $k>0.3\hmpc$. For $\Delta^{\ell=3}_{\rm kSZ}$, $\rm Cov/\widehat{Cov}$ is consistent with 1 within the statistical uncertainty at $k<0.5\hmpc$. Overall speaking, the Gaussian approximation adopted in Eq.~(\ref{eq:cov}) works very well, and the theoretical calculation captures most components of the realistic covariance measured from simulations.

Therefore, we conclude that the covariance matrix calculated by Eq.~(\ref{eq:cov}) is accurate enough for our scientific purpose in this work, and we do not resort to a larger suite of N-body simulations for a more accurate measurement of the covariance.

\section{Accuracy comparison of the non-linear models}
\label{app:accuracy}
\begin{figure*}
\begin{center}
\includegraphics[width=2\columnwidth]{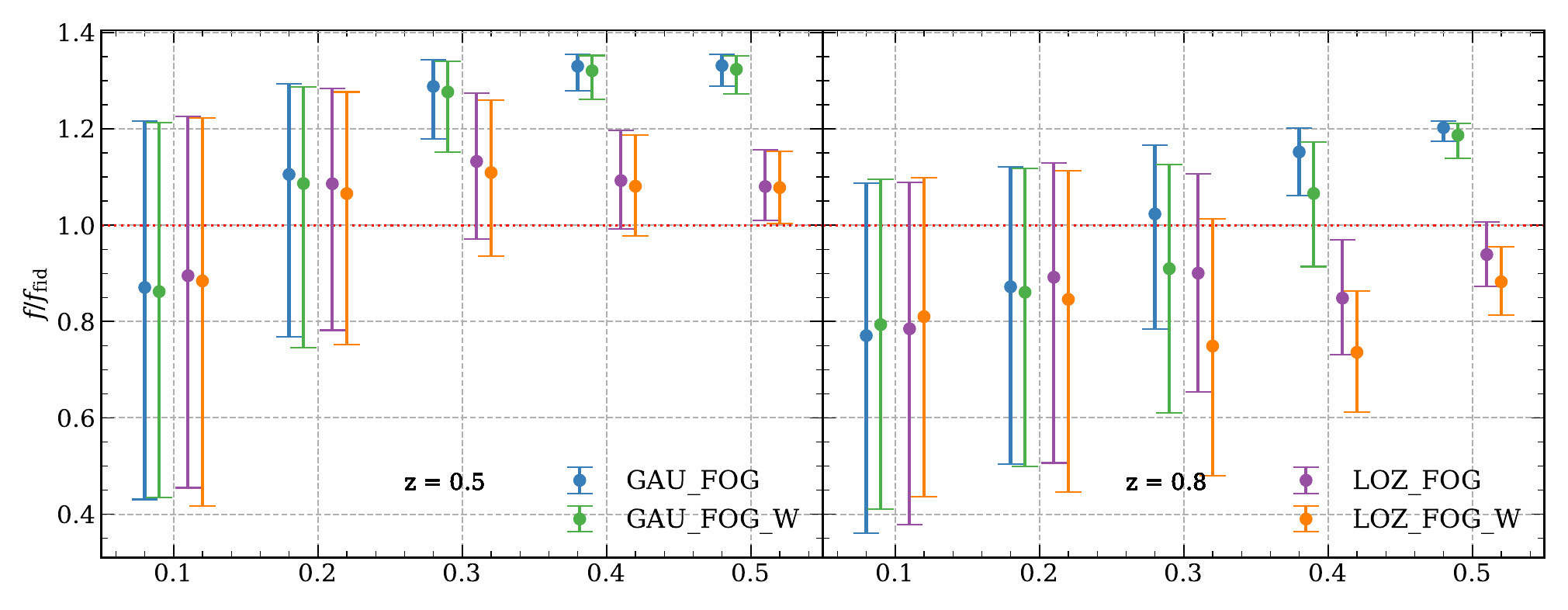}
\includegraphics[width=2\columnwidth]{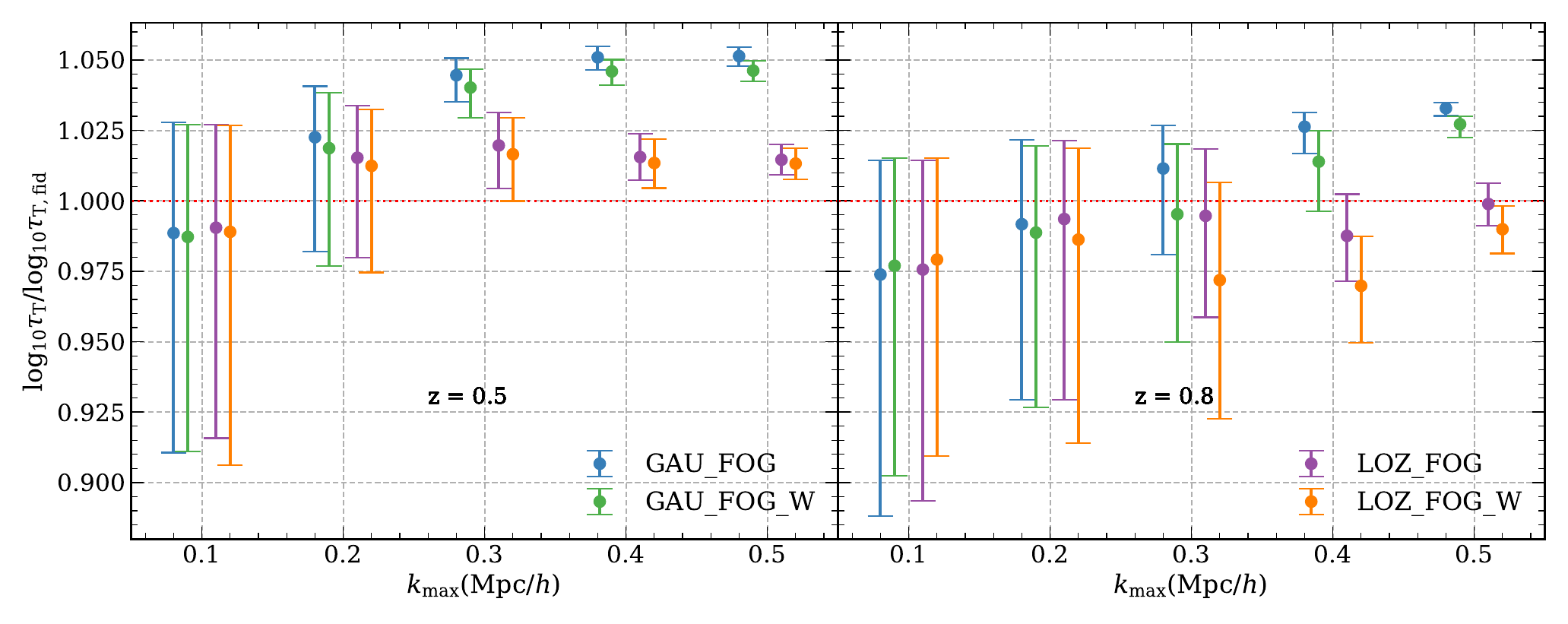}
\end{center}
\caption{The constraints on $f$ and $\log_{10}\tauT$ of 4 non-linear kSZ models from $\Delta^{\ell=1,3}_{\rm kSZ}$ of the mocks. {The models are described by Eq.~(\ref{eq:P_kSZ_Gau}) or~(\ref{eq:P_kSZ_Lor}). The notation with ``Gau/Lor" means that the Gaussian/Lorentzian FoG term is adopted and that with/without ``W" means that Eq.~(\ref{eq:W}) or $W=1$ is utilized to calculate $W(k)$.} At small scales, the small error bars of the fitted parameters from `NL-Gau' and `NL-Gau-W' models are misleading, since the fitted values are approaching the prior cut of $f=1$.}
\label{fig:accuracy}
\end{figure*}

With the MCMC technique, we compare the constraints on $f$ and $\log_{10}\tauT$ of 4 non-linear kSZ models from the measured $\Delta^{\ell=1,3}_{\rm kSZ}$ of the mock galaxies in Fig.~\ref{fig:accuracy}. In general, models with the Lorentzian FoG function behave better than those with the Gaussian one. This is understandable since our models of the Kaiser effect is simple and the residual model uncertainty is absorbed into the otherwise Gaussian FoG term~\citep{Scoccimarro04,Zheng13,Zheng16a} and make it closer to being Lorentzian. Taking a closer look at the result, the accuracy of the `NL-Lor' model is slightly better than the `NL-Lor-W' one.

Although being simple and phenomenological, the MCMC fitted parameter values of `NL-Lor' and `NL-Lor-W' models are within $1-2\sigma$ ranges of the fiducial ones. This accuracy is achieved even at non-linear scales of $k>0.3\hmpc$. It is encouraging and motivates further refinements on the current models to improve their accuracy. We leave this effort to future works. 

\section{{MCMC fitting with \texorpdfstring{\MakeLowercase{$b_1$}}{} as an additional free parameter}}
\label{app:MCMC_nonfixb1}
\begin{figure*}
\includegraphics[width=1\columnwidth]{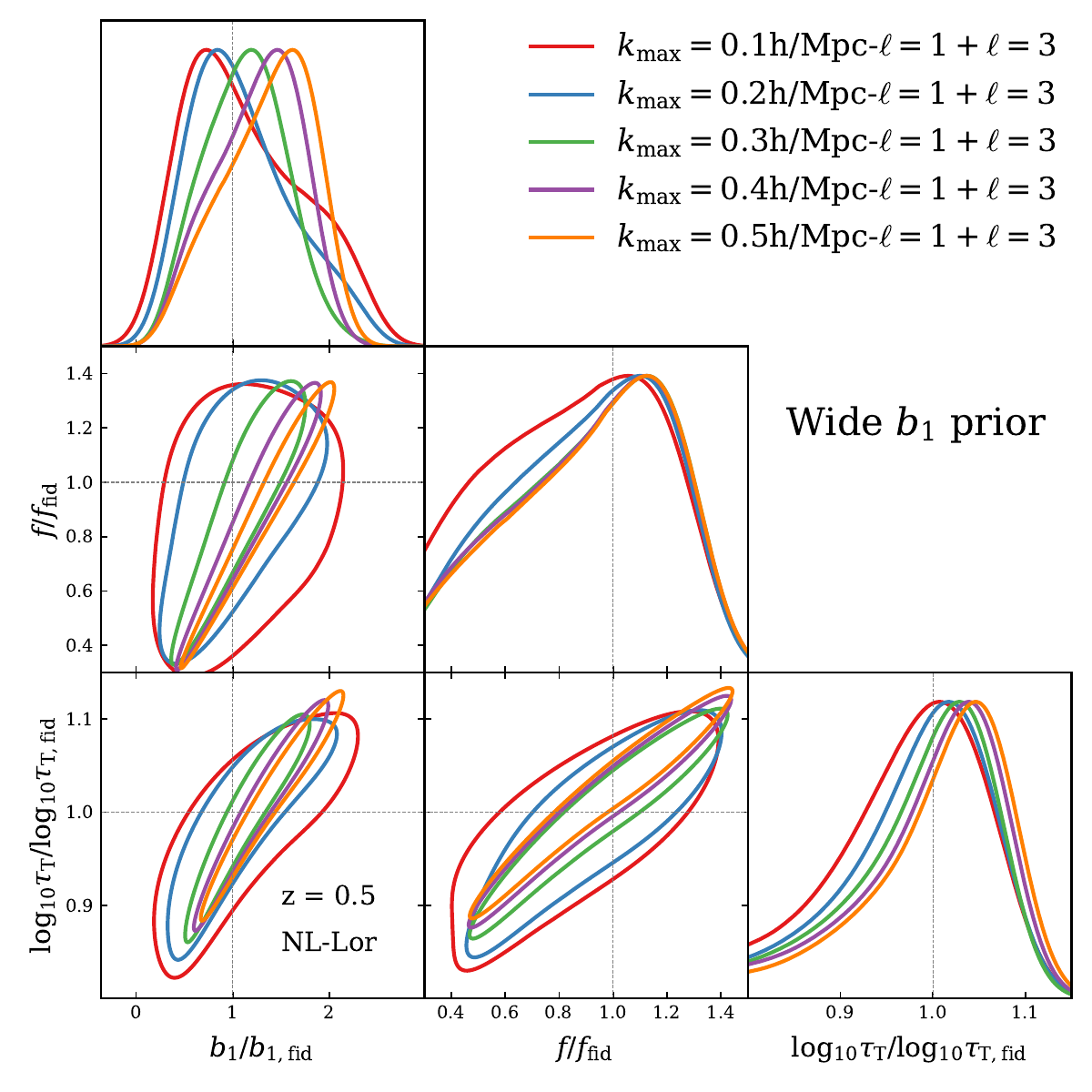}
\includegraphics[width=1\columnwidth]{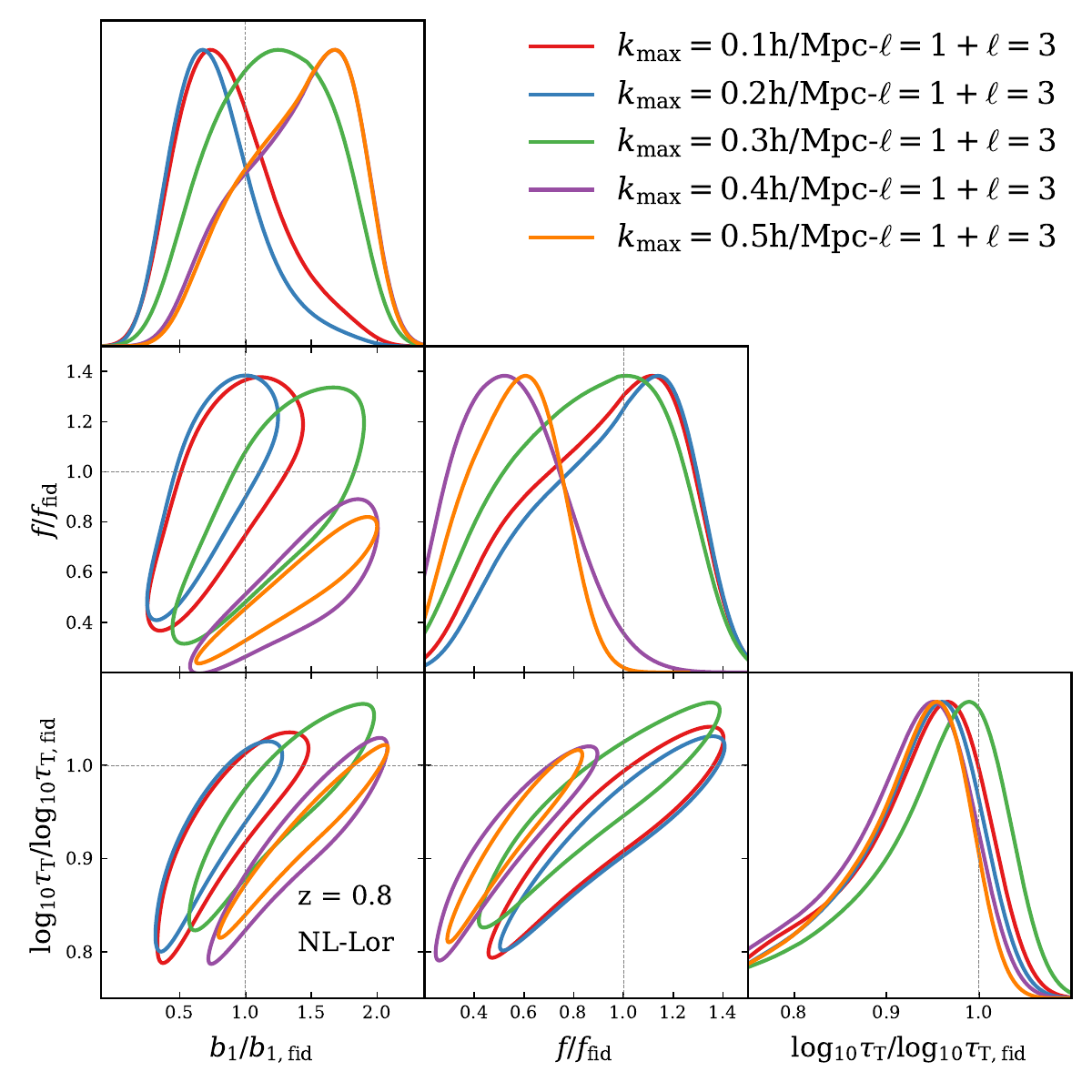}
\caption{{The same as Fig.~\ref{fig:MCMC_nonfixedbk}, except that we only include $b_1$ as an additional free parameter in the MCMC fitting.}}
\label{fig:MCMC_nonfixedb1}
\end{figure*}
{In Fig.~\ref{fig:MCMC_nonfixedb1} we present the MCMC fitting results when we adopt a linear bias model and include $b_1$ as an additional free parameter. Two things are clearly illustrated: first, as already shown in Fig.~\ref{fig:MCMC_nonfixedbk}, additional bias parameter(s) weakens the $\tauTf$ degeneracy breaking; second, the wrong (linear) bias model induces systematic errors on the fitted $f$ and $\tauT$ at $z=0.8$.}

\bsp	
\label{lastpage}
\end{document}